\documentclass[12pt]{article}
\usepackage{xcolor,enumitem,mathtools,amsmath,amsfonts,amssymb,cite,float,bbold,accents}

\usepackage[unicode,colorlinks = true,
            linkcolor = blue!70!black,
            urlcolor  = red!70!black,
            citecolor = green!55!black,
            anchorcolor = blue!70!black]{hyperref}
\usepackage[margin=0.8in]{geometry}

\DeclareMathAlphabet\mathbfcal{OMS}{cmsy}{b}{n}

\newcommand\thickbar[1]{\accentset{\rule{.5em}{.7pt}}{#1}}
\def\di{\mathrm{d}}
\def\q{\boldsymbol{q}}
\def\qalpha{\boldsymbol{q}^{(\alpha)}}
\def\qalphaT{\boldsymbol{q}^{(\alpha) \top}}
\def\qalphaprime{\boldsymbol{q}^{(\alpha')}}
\def\Tvec{\boldsymbol{T}}
\def\tauvec{\boldsymbol{\tau}}
\def\thetavec{\boldsymbol{\theta}}
\def\phivec{\boldsymbol{\phi}}
\def\thetatildevec{\widetilde{\boldsymbol{\theta}}}
\def\Qtilde{\widetilde{\mathbfcal{Q}}}
\def\Qbar{\thickbar{\mathbfcal{Q}}}
\def\Lambdatilde{\widetilde{\Lambda}}
\def\Lambdabar{\thickbar{\Lambda}}
\def\alphavec{\boldsymbol{\alpha}}
\def\qbar{\thickbar{\boldsymbol{q}}}


\begin{document}
%%%%%%%%%%%%%%%%%%%%%%%%%%%%%%%%%%%%%%%%%%%%%%%%%%%%%%%%%%%
\begin{titlepage}
\hfill \\
\vspace*{15mm}
\begin{center}
{\LARGE \bf Catastrophic Inflation in the Axiverse}
\vspace*{15mm}

{\large Naomi Gendler$^{\,\rm a}$, Oliver Janssen$^{\,\rm b}$, Matthew Kleban$^{\,\rm c}$ and Cameron Norton$^{\,\rm c}$}

\vspace*{8mm}

{\small 
$^{\rm a}$
Jefferson Physical Laboratory, Harvard University, Cambridge, MA  02138, USA
\\[.1cm]
$^{\rm b}$
Laboratory for Theoretical Fundamental Physics, EPFL, 1015 Lausanne, Switzerland
\\[.1cm]
$^{\rm c}$
Center for Cosmology and Particle Physics, New York University, New York, NY
10003, USA}

\vspace*{0.7cm}
\end{center}
\begin{abstract}

\noindent \normalsize In the landscape of compactifications of type IIB string theory on O3/O7 orientifolds of Calabi--Yau threefolds, we identify regions of the axion potential in specific geometries that support small-field inflation, under the assumption that the overall Calabi--Yau volume can be stabilized at certain tuned values.  At the tuned volume multiple critical points of the axion potential merge in a ``catastrophe'', creating a region with small gradient and small Hessian determinant --- necessary conditions for slow-roll inflation.  We describe an algorithm that identifies several explicit examples of these catastrophes that have small field range $f / M_{\rm Pl} \ll 1 $ yet inflate for thousands of efolds.  
Some examples correctly reproduce the observed spectral tilt but not the  amplitude of scalar perturbations, while others reproduce the observed amplitude but with a tilt that is slightly too red. All our examples have negligible tensor perturbations and no significant non-Gaussianity generated during the slow-roll phase.  Our results demonstrate that string theory axions can be the source of cosmic inflation without alignment or large  field ranges, provided that the K\"ahler moduli can be stabilized at special values of the volume.
\end{abstract}

\vspace{1cm}

\today

\end{titlepage}
%%%%%%%%%%%%%%%%%%%%%%%%%%%%%%%%%%%%%%%%%%%%%%%%%%%%%%%%%%%

\tableofcontents

%%%%%%%%%%%%%%%%%%%%%%%%%%%%%%%%%%%%%%%%%%%%%%%%%%%%%%%%%%%
\section{Introduction} \label{sec:intro}
%%%%%%%%%%%%%%%%%%%%%%%%%%%%%%%%%%%%%%%%%%%%%%%%%%%%%%%%%%%
Many of the properties of our universe could arise from a large landscape of
possible theories of low-energy physics
\cite{Bousso:2000xa,Susskind:2003kw,Vafa:2005ui,Douglas:2006es,Denef:2007pq}.
The compactification of extra dimensions generically gives rise to many scalar
and pseudoscalar fields in the four-dimensional effective description, including
multiple axion fields \cite{Svrcek:2006yi,Arvanitaki:2009fg}.  The potentials
for these fields are highly complex and the full landscape can have an enormous
number of local minima, as well as many other features.  This structure has been
explored as a potential solution to the cosmological constant problem, and as a
framework for addressing other macroscopic features of our universe such as
dark matter, inflation, and the strong CP problem
\cite{Weinberg:1987dv,Kachru:2003aw,Baumann:2014nda}.

In this paper we will be concerned with the question of whether slow-roll inflation can arise from the axion sector of string theory compactifications.  Previous approaches to this problem utilized some version of an alignment mechanism, typically relying on a large number $N$ of axion fields.  For instance, the result of \cite{Bachlechner:2017hsj,Bachlechner:2017zpb} was that the field range can be enhanced from $f$ to $\sim N^{3/2} f$, while the number of minima scales as $(N!)^{1/2}$.  However, these results are valid  in a regime where the charge matrix of the axion fields is not too sparse.  The work of \cite{Gendler:2023kjt} showed that
the number of minima is small in a large set of individual Calabi--Yau compactifications, which is a consequence of the sparsity of the charge matrices.  Correspondingly, the alignment mechanisms that can give rise to a field range enhanced by $N^{3/2}$ do not appear to be available in these compactifications \cite{Long:2016jvd}.

Nevertheless, we will show that even in compactifications with few local minima and small $N$, slow-roll inflation in the axion sector  can occur at  special points in the moduli space of certain compactifications where two or more critical points of the potential merge.  Such points are known as catastrophes, which were previously considered in the context of inflation in \cite{Downes:2011gi}. They are characterized by the vanishing of both the gradient and one of the eigenvalues of the Hessian.

All the examples we find are fairly well-approximated by single-field quartic hilltop or cubic shoulder models of inflation with sub-Planckian field ranges.  Small-field quartic hilltop models  predict a scalar tilt that is too red to agree with observations \cite{German:2020rpn}, while cubic shoulder models can accommodate it.  In this preliminary work we have not identified an example that agrees with both the tilt and the amplitude of the scalar power spectrum, but there is no obvious obstacle to finding one with a more complete scan.

Previous work on inflation in string theory \cite{Baumann:2014nda} includes brane inflation~\cite{Dvali:1998pa, Kachru:2003sx}, in which the inflaton is the separation between D3 and anti-D3 branes; K\"{a}hler moduli and fiber inflation~\cite{Balasubramanian:2005zx, Cicoli:2008gp}, in which the inflaton is a blow-up or fiber modulus in the Large Volume Scenario; and axion monodromy inflation~\cite{Silverstein:2008sg, McAllister:2008hb}, which exploits monodromy to generate a large field range. Natural inflation~\cite{Freese:1990rb} and its string-theoretic realizations exploit the discrete shift symmetry of axions to protect the flatness of the potential. Our work can be viewed as a small-field extension of that program.

%%%%%%%%%%%%%%%%%%%%%%%%%%%%%%%%%%%%%%%%%%%%%%%%%%%%%%%%%%%
\subsection{Axion potentials}
%%%%%%%%%%%%%%%%%%%%%%%%%%%%%%%%%%%%%%%%%%%%%%%%%%%%%%%%%%%
Axions arising from perturbative  string theory compactifications typically (or perhaps, always \cite{Vafa:2005ui,Svrcek:2006yi,Ooguri:2006in}) have sub-Planckian field ranges, $f / M_{\rm Pl} \ll 1$, where $f$ is the axion decay constant that sets the field-space range over which the axion potential is periodic.  Schematically, the axion potential consists of sums of cosines of linear combinations of the axion fields $\phi^i$, multiplied by amplitudes $\Lambda_I$ that depend exponentially on the compactification volume and hence vary over a very wide range: 
\begin{equation}
V(\phivec) \sim \sum_I \Lambda_I^4 \cos \left( \q_I \cdot \phivec/f \right) \,.
\end{equation}

For generic values of $\Lambda_I$ and $\q_I$, potentials of this form do not give rise to slow-roll inflation.  The slow-roll parameters are
\begin{equation}
    \varepsilon_V = \frac{M_{\rm Pl}^2}{2} \left( \frac{V'}{V} \right)^2  \,, \quad \eta_V = M_{\rm Pl}^2 \frac{V''}{V} \,.
\end{equation}
Dimensional analysis indicates $M_{\rm Pl}^2 (V'/V)^2 \sim M_{\rm Pl}^2 V''/V \sim M_{\rm Pl}^2/f^2$, which  satisfies $(M_{\rm Pl}/f)^2\gtrsim 10^2$  for the models we consider. Indeed, for small values of $f$ (and in general for small-field models), slow-roll inflation can only arise if the potential is tuned to create flat regions where these derivatives are much smaller than  dimensional analysis indicates.

We will demonstrate that at certain values of the overall volume of the compactification manifold it is possible to arrange for a near-cancellation of a few terms in the axion potential that contribute to the slow-roll parameters. This cancellation  creates small, nearly flat regions of an otherwise steep potential, allowing for a period of slow-roll inflation. This mechanism is distinct from  models that rely on a large number of axion fields \cite{Bachlechner:2017hsj,Bachlechner:2017zpb}, and indeed we find explicit examples with  a number of axions $h^{1,1}  < 10$.  The caveat is that achieving $\mathcal{O}(50)$ efolds of inflation requires tuning of the $\Lambda_I$ (or correspondingly, the Calabi--Yau volume) at the $(f/M_{\rm Pl})^2$ level.  Without self-consistently stabilizing the K\"ahler moduli, we cannot be sure that such tuned values of the volume are physically allowed.

We will focus our analysis on the Kreuzer--Skarke set of Calabi--Yau threefolds, a well-studied class of geometries that give rise to calculable axion potentials. In previous work, it was shown that these potentials, in a regime of perturbative control, typically have only a handful of distinct minima \cite{Gendler:2023kjt}. This is due to the sparse nature of the charge matrix and the steep exponential hierarchies in the potential's Fourier coefficients. While this result suggests that potentials corresponding to the individual geometries are probably not ``rich'' enough for an anthropic solution to the cosmological constant problem, the class of geometries is enormous, potentially allowing for a solution within this larger set.  These previous results left open the question of whether the axion sector of individual geometries can support slow-roll inflation.

%%%%%%%%%%%%%%%%%%%%%%%%%%%%%%%%%%%%%%%%%%%%%%%%%%%%%%%%%%%
\subsection{Toy model} \label{Toy model}
\begin{figure}[!ht]
    \centering
    \includegraphics[width=\textwidth]{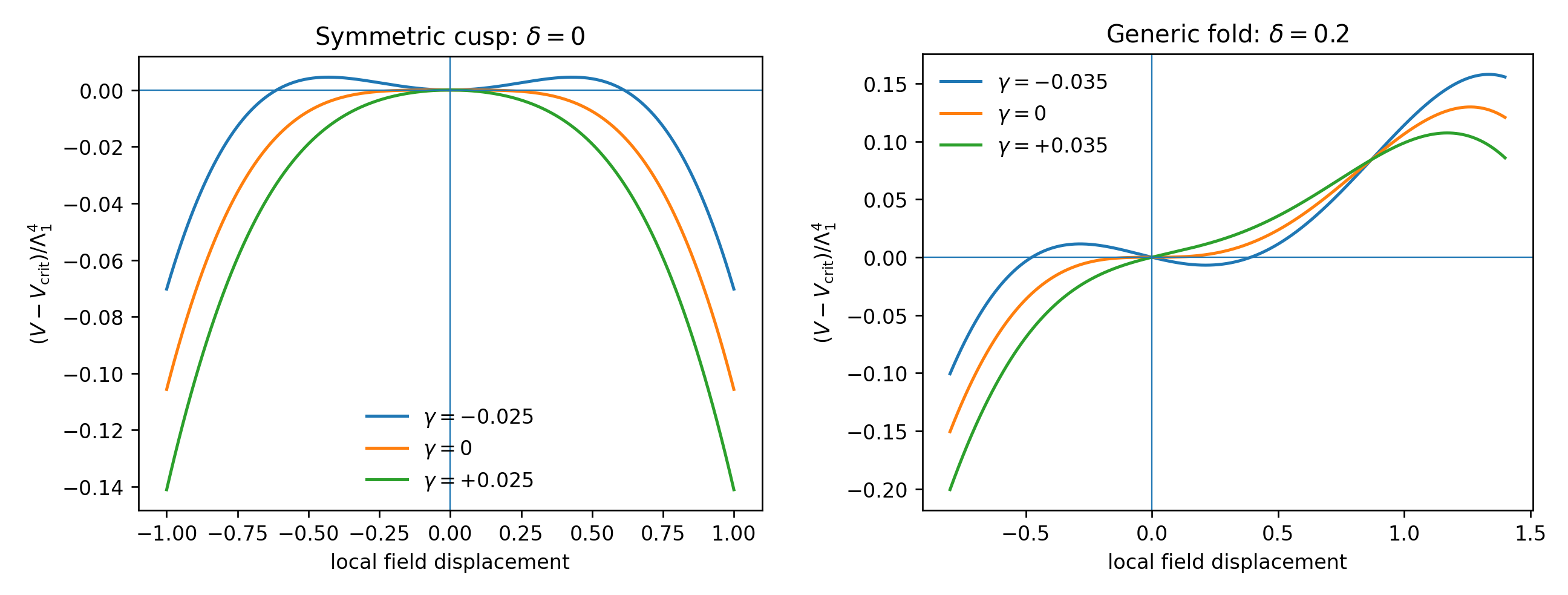}
    \caption{Catastrophes in the two-cosine model with
    $q_1=1$ and $q_2=2$.  \textbf{Left}: for vanishing relative phase
    $\delta=0$, the reflection-symmetric potential with
    $a \equiv \Lambda_2^4/\Lambda_1^4 = a_c - \gamma$ has a cusp at $a = a_c = 1/4$, where a local minimum and two neighboring maxima
    merge into a quartic hilltop.  \textbf{Right}: the symmetry is broken for  
    nonzero phase, here $\delta = 0.2$.  With $a = a_c(\delta) - \gamma$, tuning $\gamma = 0$ produces a fold where a local
    maximum and minimum merge into a cubic shoulder. }
    \label{fig:toy_models}
\end{figure}

%%%%%%%%%%%%%%%%%%%%%%%%%%%%%%%%%%%%%%%%%%%%%%%%%%%%%%%%%%%
To understand the high-dimensional catastrophes that make inflation in the
axion sector possible, it is useful to begin with a single-field toy model:
\begin{equation} \label{toyaxionV}
V(\theta)
=
\Lambda_1^4\left[1-\cos q_1\theta\right]
+
\Lambda_2^4\left[1-\cos\left(q_2\theta+\delta\right)\right] \,.
\end{equation}
  Here $q_1, q_2$ are integers, $\delta$ is the relative phase, and $\theta$ is a dimensionless axion angle; locally the canonically normalized
field is $f \theta$.  
For $\delta = 0$ this potential is minimized at $\theta = 0$, while at $\theta = \pi$,
\begin{equation}
V'(\pi) = 0 \,, \quad V''(\pi ) =  q_1^2 \Lambda_1^4  \cos{q_1 \pi} +  q_2^2\Lambda_2^4 \cos{q_2 \pi} \,. \quad \quad \text{(when } \delta = 0 \text{)}
\end{equation}
The second derivative can be tuned to small values.  For this to occur, the higher-frequency (larger $q$) term should have a smaller coefficient than the lower-frequency one.\footnote{This aligns with the structure of the  axion potential in string theory,  where $\Lambda^4\sim \exp(- 2 \pi q \tau)$, where $q$ is the charge vector and $\tau$ is the volume of the corresponding cycle,  scaling as the overall volume of the Calabi--Yau.  When the volume is large, terms with different $q$ will have very large ratios and the potential is dominated by a few large terms.  At small volume corrections are large and the form of the potential cannot be trusted.  As we will see, there is an intermediate regime where things can be interesting.}
For instance, if $q_1 = 1, q_2 = 2,$ and $\Lambda_2^4 = \Lambda_1^4/4$, then $V''(\pi) = 0$.  Here the vanishing of the second derivative at $\theta = \pi $ arises due to the collision of three critical points: two local maxima and one local minimum.  To see this, consider $a \equiv \Lambda_2^4/\Lambda_1^4 = 1/4 - \gamma$.  As $\gamma$ is increased to zero, the three critical points merge, and at $\gamma = 0$ there is a single, quartic maximum (Fig.~\ref{fig:toy_models}, left panel).  This is known as a ``cusp'' catastrophe.  

For $\delta=0$, the potential is reflection-symmetric about $\theta=\pi$.
At $a_c=1/4$, not only $V'$ and $V''$ but also $V'''$ vanish.  Writing $x\equiv\theta-\pi$ and using
$a=1/4-\gamma$, one finds
\begin{equation}
V(x)
=
2\Lambda_1^4\left[
1-\gamma x^2
-\left(\frac1{16}-\frac{\gamma}{3}\right)x^4
+\mathcal O(x^6)
\right] \,.
\label{eq:symmetric-two-cosine-expansion}
\end{equation}
  For $\gamma<0$ the central point is a local
minimum flanked by two maxima, whereas for $\gamma>0$ it is a hilltop.

At generic  $\delta \neq 0$, the reflection symmetry is broken.  Nevertheless, a ``fold'' catastrophe occurs when two critical points merge (Fig.~\ref{fig:toy_models}, right panel).  The fold occurs at $\delta$-dependent values of $\theta$ and $a_c(\delta)$.
At the fold $\Lambda_2^4/\Lambda_1^4 \equiv a = a_c$ the leading
non-constant term is cubic, while detuning $a = a_c - \gamma$ produces a linear term for $\gamma \neq 0$.  More explicitly, let $\theta_{\rm inf}$ be the
 point near the detuned fold at which $V''=0$ and set
$y\propto \theta-\theta_{\rm inf}$. 
The potential becomes
\begin{equation}
V(y)
=
V_0\left(
1-\lambda y
- \frac{c_3}{3} y^3
+\cdots
\right) \,,
\qquad
\lambda\propto\gamma \,.
\label{eq:cubic-linear-normal-form}
\end{equation}
Exactly on the fold, $\gamma=\lambda=0$, so the local potential is a constant
plus a cubic.  Detuning the amplitude introduces the linear term  (the
quadratic term remains absent by construction, because the expansion is  about the 
inflection point).

Because both the first and second derivatives vanish, both the quartic cusp and the cubic fold can support many efolds of inflation despite a
sub-Planckian field range.  For later comparison with numerical trajectories, write
the quartic potential as
\begin{equation}
V(x)
=
V_0\left(
1-\frac{\beta}{2}x^2-\frac{c_4}{4}x^4+\cdots
\right) \,,
\qquad
x\equiv \theta - \pi \,,
\label{eq:quartic-toy-normalized}
\end{equation}
where $c_4>0$.  For the symmetric two-cosine model above,
$\beta=2\gamma+\mathcal O(\gamma^2)$ and
$c_4=1/4+\mathcal O(\gamma)$.
In our string theory examples the control parameter is
the compactification-volume parameter $k$; near the catastrophe,
\begin{equation}
\beta(k)=\beta_1\Delta k \,,
\qquad
\Delta k\equiv k-k_c \,,
\label{eq:beta-deltak}
\end{equation}
with $\beta_1>0$.  
 Approximating $V\simeq V_0$ in the numerator and defining a canonically normalized field $\phi = f x$, the number of efolds is
\begin{align}
\mathcal{N}_e
&\simeq
\frac{1}{M_{\rm Pl}^2}
\int_{\phi_{\rm in}}^{\phi_{\rm end}} |\di\phi| \,  
\frac{V}{V'}
\nonumber\\
&\simeq
\frac{f^2}{2\beta M_{\rm Pl}^2}
\log\left(
\frac{x_{\rm end}^2\left(\beta+c_4 x_{\rm in}^2\right)}
{x_{\rm in}^2\left(\beta+c_4 x_{\rm end}^2\right)}
\right) \,. \label{eq:efolds-quartic-toy}
\end{align}
The end of slow-roll is approximately determined by $|\eta_V|\simeq1$, with
\begin{equation}
\eta_V
\simeq
-\frac{M_{\rm Pl}^2}{f^2}\left(\beta+3c_4 x^2\right) \,,
\end{equation}
so that
\begin{equation}
x_{\rm end}^2
\simeq
\frac{f^2/M_{\rm Pl}^2-\beta}{3c_4} \,,
\label{eq:x-end-quartic}
\end{equation}
provided $\beta<f^2/M_{\rm Pl}^2$.  In the regime
$c_4 x_{\rm end}^2\gg\beta$, the efold count simplifies to
\begin{equation}
\mathcal{N}_e
\simeq
\frac{f^2}{2\beta M_{\rm Pl}^2}
\log\left(
1+\frac{\beta}{c_4 x_{\rm in}^2}
\right) \,.
\label{eq:efolds-quartic-simple}
\end{equation}
Using \eqref{eq:beta-deltak}, this becomes
\begin{equation}
\mathcal{N}_e
\simeq
\frac{f^2}{2\beta_1\Delta k\,M_{\rm Pl}^2}
\log\left(
1+\frac{\beta_1\Delta k}{c_4 x_{\rm in}^2}
\right) \,.
\label{eq:efolds-quartic-deltak}
\end{equation}
This scaling explains why tuning $\Delta k$ and $x_{\rm in}$ close to zero produces
many efolds even when the available field range is small.\footnote{As with any
hilltop model, the $x_{\rm in}\to0$ divergence is regulated by de Sitter fluctuations:
the initial displacement should not be taken smaller than
$\delta\phi\sim H$, or $x_{\rm in}\gtrsim H/f$.  In the pure quartic
limit this gives $\mathcal N_{e,\max}\sim f^4/V_0$ up to numerical factors.}  The cubic fold case exhibits the same qualitative features.
Appendix~\ref{infparams} gives the slow-roll observables for both the quartic
and cubic normal forms.

This mechanism generalizes to higher-dimensional field spaces.  Consider a
local minimum and one or more nearby saddles of index one (all but one Hessian eigenvalue positive).  If the potential
can be tuned so that these critical points merge, the resulting critical point
has vanishing gradient and one vanishing Hessian eigenvalue, with the
rest of the eigenvalues positive.  This produces a parametrically flat
direction that can support slow-roll inflation.  More generally, catastrophes involving higher index saddles can
have additional mildly tachyonic directions and still allow inflation; provided the corresponding
$|\eta_V|$ values are small, these fields evolve slowly and behave as light
spectators during inflation.

In the rest of this paper we use this intuition to find very flat regions of
the full axion potential.

%%%%%%%%%%%%%%%%%%%%%%%%%%%%%%%%%%%%%%%%%%%%%%%%%%%%%%%%%%%
\section{Axion potentials in string theory} \label{stringsec}
%%%%%%%%%%%%%%%%%%%%%%%%%%%%%%%%%%%%%%%%%%%%%%%%%%%%%%%%%%%
The Kreuzer--Skarke (KS) axiverse is the collection of axion sectors arising in type IIB compactifications on O3/O7 orientifolds of Calabi--Yau threefolds  constructed as hypersurfaces in toric varieties defined by triangulations of the four-dimensional reflexive polytopes in the KS database \cite{Kreuzer:2000xy,Batyrev}.  Specifically, a fine, regular, star triangulation of a reflexive polytope $\Delta^\circ$ determines a toric fourfold $V$, and a generic anticanonical hypersurface $X\subset V$ is a Calabi--Yau threefold.  Different polytopes, and in general different triangulations of the same polytope, carry different topological data and therefore different four-dimensional axion sectors.  The integer $N=h^{1,1}(X)$ counts the K\"ahler moduli of $X$, and hence the number of axions of the type considered here.\footnote{We do not construct explicit orientifolds: rather, we assume that $h^{1,1}_{-}=0$, so that no K\"ahler moduli are projected out of the effective theory.} The full KS list contains examples with $h^{1,1}$ as large as $491$.

The relevant chiral multiplets are the complexified K\"ahler moduli
\begin{equation}
    T^i = \tau^i + i\theta^i, \qquad i=1,\ldots,N \,.
\end{equation}
The saxions $\tau^i$ measure volumes of holomorphic four-cycles, or divisors, while the axions $\theta^i$ arise by reducing the Ramond--Ramond four-form $C_4$ on the same divisors,
\begin{equation}
    \tau^i = \frac12\int_{D_i} J\wedge J, \qquad
    \theta^i = \int_{D_i} C_4 \,.
\end{equation}
The continuous shift symmetry of $\theta^i$ is inherited from the gauge symmetry of $C_4$ and is broken only by non-perturbative effects.  The K\"ahler form $J$ must lie in the K\"ahler cone of $X$:
\begin{align}
    \mathcal{K}_X = \bigg\{ J ~ \bigg| \, \int_{C_i} J > 0 \bigg\} \,,
\end{align}
where the $C_i$ are the calibrated two-cycles in $X$.  Its dual cone, the Mori cone, is generated by effective curve classes, while the effective cone of divisors determines which four-cycles are holomorphic\footnote{Not every lattice point in the effective cone of divisors hosts a holomorphic four-cycle \cite{Gendler:2026uux}.}  and can therefore potentially support Euclidean D3-brane instanton corrections to the superpotential.\footnote{Whether a given D3-brane instanton actually contributes a non-zero term to the superpotential depends on the fermionic zero modes that it carries  \cite{Witten:1996bn} --- we will not treat this here. Rather, we will assume that every class in the semi-group generated by prime toric divisors gives rise to a non-trivial superpotential correction.} In practice, we will use toric data to approximate the Mori cone and the effective cone by divisors inherited from the ambient toric variety.  In favorable geometries there are $h^{1,1}+4$ prime toric divisors, which provide a sparse set of candidate instanton charges.

We will not undertake the task of moduli stabilization in this work. The examples in \S\ref{resultssec} have fixed values of the saxions, but it remains to be seen whether the fields can be stabilized dynamically at those points. After taking the saxions to have fixed values, we study the resulting effective theory for the axions. At tree level the K\"ahler potential and superpotential take the form
\begin{align}
    \mathcal{K} &= -2\log (\mathcal{V} +\mathcal{V}_{\mathrm{corr}}), \label{KahlerpotentialEq} \\
    W &= W_0 + \sum_\alpha A_\alpha \exp\left(-2\pi \qalpha\cdot\Tvec\right) \,, \label{superpotentialEq}
\end{align}
where $\mathcal{V}$ is the volume of $X$, $\mathcal{V}_{\mathrm{corr}}$ encapsulates corrections to the K\"ahler potential, $W_0$ is the flux superpotential, $A_\alpha$ are one-loop determinants, and $\qalpha$ is the integer divisor charge of the instanton.  Since the coefficients scale as $\exp(-2\pi\qalpha\cdot\tauvec)$, small changes in cycle volumes produce large hierarchies among the Fourier components of the axion potential.  This hierarchy is central to the earlier observation that controlled KS axion potentials usually have only a few distinct minima \cite{Gendler:2023kjt}, and it is also central to the present mechanism: slow-roll regions can occur only where several otherwise steep terms are tuned to compete and nearly cancel.

The potential, $V(\thetavec)$, can be expressed as a sum over instanton contributions \cite{Demirtas:2018akl}:
\begin{align} \label{Vfull}
&V(\thetavec) = -\frac{8 \pi}{\mathcal{V}^2} \left[ \sum_{\alpha} (\qalpha \cdot \tauvec) \, e^{-2 \pi \qalpha \cdot \tauvec} \cos \left(2 \pi \qalpha \cdot \thetavec + \delta_\alpha \right) + \right. \\ &\frac{1}{2}
\left.\sum_{\alpha \neq \alpha'}\left( \pi \, \qalphaT \boldsymbol{K}^{-1} \qalphaprime + \left( \qalpha + \qalphaprime \right) \cdot \tauvec \right) e^{-2 \pi \left( \qalpha + \qalphaprime \right) \cdot \tauvec} \cos \left( 2 \pi \left( \qalpha - \qalphaprime \right) \cdot \thetavec + \delta_{\alpha, \alpha'} \right) \right] \,, \nonumber 
\end{align}
where the charges $\qalpha$ are integer vectors corresponding to effective divisors in the Calabi--Yau geometry, and the phases are determined by the phases of $W_0$ and $A_\alpha$.  The $\theta^i$ are dimensionless fields with period one and a non-canonical kinetic term.  The coefficients of the cosines, proportional to $(\qalpha \cdot \tauvec) \, e^{-2 \pi \qalpha \cdot \tauvec}$, are exponentially suppressed by the volumes of the corresponding cycles.   As we will see, attaining $\mathcal{O}(50)$ efolds of inflation requires tuning the volume to roughly one part in $10^4$, so the assumption of saxion stabilization at the tuned point is non-trivial. 

Given the steep exponential hierarchies in \eqref{Vfull}, the  axion potential can be accurately approximated by a truncated sum of its most dominant terms \cite{Bachlechner:2017hsj}:
\begin{align} \label{VP}
V(\thetavec) \approx V_P(\thetavec) =  - \sum_{I=1}^P \Lambda_I^4 \, \cos \left(2 \pi (\mathbfcal{Q} \thetavec)^I + \delta_I \right) \,.
\end{align}
In the large volume limit these leading terms dominate, typically leading to a simple potential with a single or only a few minima.   At smaller volumes more terms become significant, leading to potentials with a more complicated structure.

%%%%%%%%%%%%%%%%%%%%%%%%%%%%%%%%%%%%%%%%%%%%%%%%%%%%%%%%%%%
\paragraph{Perturbative control}
%%%%%%%%%%%%%%%%%%%%%%%%%%%%%%%%%%%%%%%%%%%%%%%%%%%%%%%%%%%
In the effective theories reported on in \S\ref{resultssec}, known corrections to \eqref{Vfull} are small. To ensure this,  we require that divisor volumes and curve volumes are large enough such that analysis of \eqref{Vfull} is robust against additional contributions. These corrections come in two classes:

\begin{enumerate}
    \item Corrections to the superpotential:  \eqref{superpotentialEq} is an infinite sum of instanton contributions. To ensure convergence of this expansion, we ensure that all prime toric divisors in the compactification geometry have volumes larger than $1$ in string units. In addition to prime toric divisors, which descend from holomorphic divisors in the ambient variety, there can exist so-called \textit{autochthonous} divisors which are holomorphic only at the level of the Calabi--Yau \cite{Demirtas:2018akl}. In the present work we neglect potential contributions from instantons wrapping such divisors.
    \item Corrections to the K\"ahler potential: the  K\"ahler potential given in \eqref{KahlerpotentialEq} can receive perturbative corrections in the $\alpha'$ expansion in addition to non-perturbative corrections from worldsheet instantons. To leading order, worldsheet instanton corrections take the schematic form
    \begin{align}
        \mathcal{V}_{\mathrm{corr}} \propto \sum_{\mathbb{q} \in \mathcal{M} } \mathcal{N}_{\mathbb{q}} e^{-2\pi \mathbb{q} \cdot \boldsymbol{t}}
    \end{align}
    where $\mathbb{q}$ are the charges of holomorphic curves in $X$, living in the Mori cone $\mathcal{M}$. The $\boldsymbol{t}$ are the volumes of a basis of these curves. $\mathcal{N}_\mathbb{q}$ are the Gopakumar--Vafa (GV) invariants of each curve. Crucially, the $\mathcal{N}_\mathbb{q}$ typically grow exponentially. 

    Lattice rays in $\mathcal{M}$ can be classified as \textit{potent} or \textit{nilpotent}. Potent rays are composed of an infinite series of lattice sites with non-vanishing GV invariants, while nilpotent rays host only a finite set of states with non-vanishing GV invariants. As a proxy for control of the K\"ahler potential approximation used here, we will ensure that all curves living on potent rays have volumes greater than $1$. See \cite{Demirtas:2021nlu, Gendler:2022ztv} for a full discussion of this classification.
\end{enumerate}
It should be noted that there may well be additional moduli-dependent corrections to the effective theory other than those listed above, leaving open the question of whether our examples are truly under control. In general, given that the potential simplifies drastically at large volume and
our mechanism requires several terms to interfere, it is to be expected that our examples in \S\ref{resultssec} should
lie in the relatively small volume regime where corrections can be significant.

Often, a slightly stronger condition is imposed on the K\"ahler moduli to ensure control of the effective theory: that the moduli are constrained to live in the stretched K\"ahler cone, defined by requiring \textit{all} effective curve volumes to be at least order one in string units.  For large $h^{1,1}$, the K\"ahler cones of KS hypersurfaces are typically narrow: imposing the stretched-cone condition pushes $J$ far from the origin, makes many divisor volumes large, and strongly suppresses many instanton terms.  This is the geometric origin of the KS axiverse results of \cite{Gendler:2023kjt}: compactifications with many K\"ahler moduli naturally contain many axions with small kinetic eigenvalues, sub-Planckian geometric field ranges, and in many cases extremely small masses. The weaker, but more precise, condition that we impose here leads to qualitatively similar results.

\begin{figure}[!ht]
    \centering
    \includegraphics[width = 0.93\textwidth]{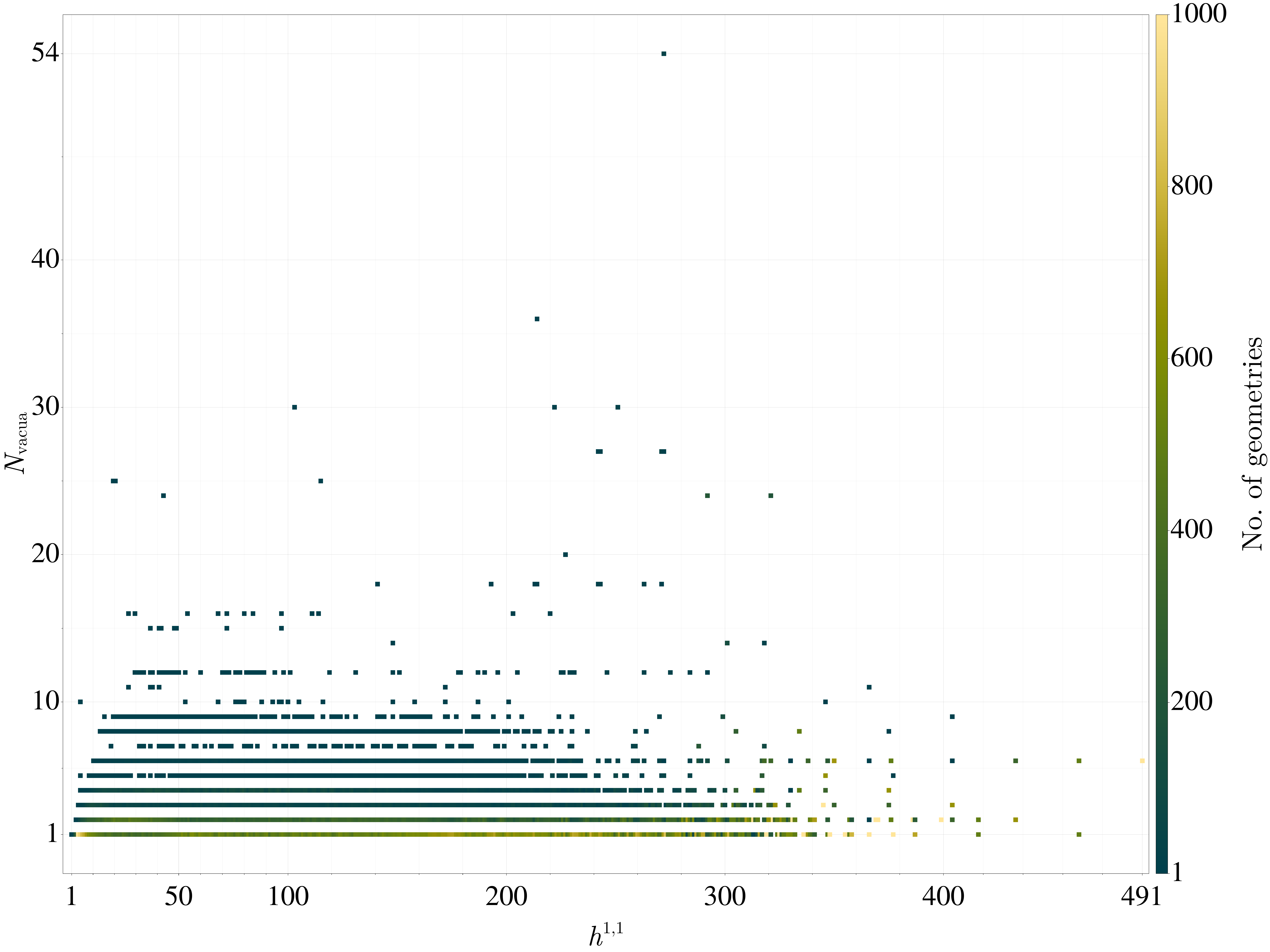}
    \caption{Distribution of numbers of distinct vacua in $\approx 400,000$ geometries versus axion number $N = h^{1,1}$. The data shows that the  majority of geometries have only a few minima, consistent with a simple potential \eqref{VP} dominated by a few terms. Figure from  \cite{Gendler:2023kjt}.}
    \label{fig:minimafig}
\end{figure}

%%%%%%%%%%%%%%%%%%%%%%%%%%%%%%%%%%%%%%%%%%%%%%%%%%%%%%%%%%%
\paragraph{Kinetic terms}
%%%%%%%%%%%%%%%%%%%%%%%%%%%%%%%%%%%%%%%%%%%%%%%%%%%%%%%%%%%
To search for slow-roll inflation, we must include the kinetic terms for the axions.  These  take the form
\begin{equation}
    \mathcal{L}_{\rm kin} = \frac{1}{2} K_{ij}\, \partial_\mu \theta^i \partial^\mu \theta^j \,,
\end{equation}
where $K_{ij}$ is the field-space metric derived from the K\"{a}hler potential:
\begin{align}
    K_{ij} = \frac{1}{2} \frac{\partial}{\partial\tau^i} \frac{\partial}{\partial \tau^j} \mathcal{K} \,.
\end{align}

The kinetic terms are relevant for determining the slow-roll parameters along an inflationary trajectory, and hence the number of efolds and inflationary observables. We stress that in this work we compute these quantities directly by working in the basis where the axion fields are canonically normalized.

The canonically normalized field is $\phi \sim K^{1/2}\theta$. For the examples given in \S\ref{resultssec}, the eigenvalues 
of $K_{ij}$ are typically of order few $\times 10^{-4}$ in Planck units.  With the potential taking the form $V \sim \Lambda^4 \cos(2\pi q \cdot \theta)$, the gradient with respect to
the canonically normalized field is $\partial V/\partial \phi \sim \Lambda^4 q/K^{1/2}$, so the slow-roll parameters
at a generic point scale as
\begin{equation}
    \varepsilon_V,\, |\eta_V| \sim \frac{q^2 M_{\rm Pl}^2}{K}  \gtrsim 10^2 \,.
\end{equation}
Slow-roll inflation is therefore impossible at a generic field-space point. The mechanism presented in this work relies on a careful cancellation of terms in $V$ and its derivatives, such that the true slow-roll parameters can be made small.

A comment is in order about the difficulty of achieving this tuning across the KS axiverse: it has been established \cite{Mehta:2021pwf, Demirtas:2018akl} that at the tip of the stretched K\"ahler cone, eigenvalues of the K\"ahler metric $K_{ij}$ tend to decrease as $h^{1,1}$ increases. This means that the catastrophe mechanism described here gets parametrically harder to achieve as $h^{1,1}$ increases.

%%%%%%%%%%%%%%%%%%%%%%%%%%%%%%%%%%%%%%%%%%%%%%%%%%%%%%%%%%%
\section{Finding inflationary regions} \label{inflationsec}
%%%%%%%%%%%%%%%%%%%%%%%%%%%%%%%%%
%%%%%%%%%%%%%%%%%%%%%%%%%%
Our goal is to find special points in the axion field space where the potential is much flatter than at a typical point, so that small-field slow-roll inflation can occur despite the small axion field range $f \ll M_{\rm Pl}$. As motivated in  \S\ref{Toy model}, this search can be phrased in the language of catastrophe theory. A generic scalar potential is a Morse function, meaning that its critical points are isolated with no zero Hessian eigenvalues. As the parameters of the potential are varied, the number of critical points remains stable except at special parameter values where critical points degenerate. At these points, at least one eigenvalue of the Hessian vanishes. 

In our problem the control parameters are the divisor volumes, which set the relative sizes of the instanton terms. We probe this parameter space along a one-dimensional trajectory defined by uniformly scaling the divisor volumes 
\begin{equation} \label{TauScaling}
    \tau^i(k) = k\,\tau^i(k=1) \,,
\end{equation}
where $\tau^i(k=1)$ are the volumes at the tip of the stretched K\"ahler cone.  Varying $k$ traces a path through the space of axion potentials. If the number of critical points changes along this path, the path must have crossed a catastrophe locus where two or more critical points merged and annihilated.

Previous work \cite{Gendler:2023kjt} developed a numerical method to count the number of critical points in these geometries.  Running this analysis at small and large $k$ to see if the number changes provides a practical diagnostic for catastrophes. Schematically, our  algorithm is as follows:
\begin{itemize}
    \item We choose a geometry from the KS database \cite{Kreuzer:2000xy}.
    \item We pick a small-volume point in the K\"ahler cone (this point can be below the tip of the stretched K\"ahler cone, out of perturbative control) and compute the coefficients $\Lambda_I^4$ for a truncated potential \eqref{VP}.  We then numerically compute the number of distinct critical points using the algorithm of \cite{Gendler:2023kjt}, reviewed in Appendix~\ref{sec:critical_points_algorithm}.
    \item We increase $k$ (the overall volume) to a larger value within the regime of perturbative control (i.e. to where all divisor volumes are bigger than or equal to $1$, and all potent curves have volumes greater than or equal to 1 --- see \S\ref{stringsec}), and  again  numerically compute the number of distinct critical points.  If this number is smaller, we conclude that critical points have annihilated in a catastrophe at some intermediate value of the overall Calabi--Yau volume.
    \item We identify the catastrophe point accurately in axion field space $\thetavec$ and in $k$.  
    We do this by solving the critical-point equations for $V(\thetavec;k)$ and tuning $k$ until the Hessian at that point develops a zero eigenvalue.  
    \item Having found a catastrophe, we  verify that the volume is sufficiently large that we are in a regime of computational control.  
\end{itemize}

To determine the inflationary dynamics, we numerically evolve the canonically normalized axion fields through the nearly flat region of the potential around the catastrophe. We choose an initial point in field space that is slightly displaced from the catastrophe along the flat direction (the direction of the Hessian eigenvector whose eigenvalue vanishes at $k = k_c$).
We work in the slow-roll approximation, where Hubble friction suppresses their acceleration and the multifield equations of motion reduce to 
\begin{equation}
    \frac{\di\phi^i}{\di\mathcal N}
    \simeq
    -\frac{M_{\rm Pl}^2}{V}
    \frac{\partial V}{\partial\phi^i} \,,
\end{equation}
with $\mathcal N$ the number of efolds. The fields therefore follow the gradient flow of the potential. We then check the consistency of this approximation along each trajectory by monitoring $\varepsilon$, $\eta_\parallel$, and the transverse Hessian eigenvalues. We count only the interval over which the slow-roll conditions are satisfied, taking inflation to end when either $\varepsilon$ or $|\eta_\parallel|$ reaches unity. Numerically, we solve the slow-roll gradient-flow equations as a continuous first-order ODE system using an adaptive stiff/non-stiff integration scheme. 

For clarity,  we summarize the set of assumptions used in obtaining the examples reported on in \S\ref{resultssec}:
\begin{enumerate}
    \item We assume that for each example, a consistent orientifold with $h^{1,1}_{-}=0$ can be found. We do not explicitly construct such orientifolds as this would not affect the overall mechanism, though it may change the nature of the specific examples.
    \item At the chosen point in moduli space (in particular, the value of $k$), the saxion fields are heavy (non-dynamical) and as such are set to their VEVs in the effective theory. Similarly, the complex structure moduli and axio-dilaton are assumed to be stabilized with large masses.
    \item We take the phases in  \eqref{Vfull} to be adjustable parameters, and we set $|W_0| = |A_\alpha| = 1$.
    \item The only contributions to \eqref{superpotentialEq} are Euclidean D3-branes wrapped on prime toric divisors and their non-negative linear combinations.
    \item We assume that the full solution is approximately de Sitter, with the global minimum of the axion potential set to $V=0$. 
    \item Unknown corrections to the K\"ahler potential \eqref{KahlerpotentialEq} are negligible.
    \item We are free to choose the initial conditions that start inflation. We do not address the measure on initial conditions.
\end{enumerate}

\newpage
%%%%%%%%%%%%%%%%%%%%%%%%%%%%%%%%%%%%%%%%%%%%%%%%%%%%%%%%%%%
\section{Explicit examples} \label{resultssec}
%%%%%%%%%%%%%%%%%%%%%%%%%%%%%%%%%%%%%%%%%%%%%%%%%%%%%%%%%%%
Our goal is to find examples of axion potentials for which the moduli can be tuned such that an inflationary region is supported. To this end, we have performed a preliminary scan over a subset of the geometries in the KS database with relatively small $N$.  As our previous results indicated, most geometries admit one or only a few minima (Fig.~\ref{fig:minimafig}).  A small fraction satisfy the criterion that the number of minima changes with increasing volume, indicating that at least one minimum has annihilated with another critical point.\footnote{For simplicity we focus on examples where the number of minima changes, because these are guaranteed not to have negative directions that could destabilize inflation.  A more complete survey would include catastrophes where saddles of higher index merge.}  Of those, we have found 24 examples which exhibit this behavior at a volume that we expect to be under reasonably good perturbative control. In this section we present two representative examples, one with $N = 5$ axions and one with $N = 8$.

Although both examples arise from multidimensional axion potentials, to a very good approximation the dynamics of the $N=5$ example can be  reduced to a single field direction in the five-dimensional axion field space, so the catastrophe is described by a single-field  potential. In the $N = 8$ example, there is no analogous simple reduction to a single field direction. Nevertheless, as we will see, the resulting trajectory only bends slightly in field space so the dynamics are effectively single-field, especially over the observable window. 

This effective one-dimensionality makes the toy models of  \S\ref{Toy model} a useful guide to the local structure of the potential near the catastrophe. In a generic one-parameter family of one-dimensional potentials, varying a single control parameter would produce a fold catastrophe: two critical points merge and annihilate (Fig.~\ref{fig:toy_models}, right panel). However, in both examples presented here, when the phases in \eqref{truncpot} are set to zero the local potential exhibits a quartic cusp, in which three nearby critical points coalesce and the leading nontrivial behavior near the hilltop is quartic rather than cubic (Fig.~\ref{fig:toy_models}, left panel). This occurs for the same reason it does in the toy model, where a symmetry is broken by a non-zero phase.

In all of our examples, once the field has evolved far enough from the flat region, inflation ends.  The axions then roll down a steep potential slope to the global minimum of \eqref{truncpot} where reheating could occur.  Because we have included a constant to set the minimum to zero vacuum energy, the scale of inflation and the amplitude of density perturbations are entirely determined, with no free parameters.

The purpose of the examples below is to demonstrate that catastrophe-induced slow-roll regions can arise in explicit KS axion potentials, rather than to provide a model that precisely matches observation.  Indeed, we have yet to find an example which agrees with both the observed values of $\delta \rho/\rho$ and $n_s$. 
If we set the phases to zero all the examples we have identified are of the small-field quartic cusp type, so that their inflationary parameters are closely related.  The maximum number of efolds is arbitrarily large (classically), and scales as $f^2/(\Delta k M_{\rm Pl}^2)$ \eqref{eq:efolds-quartic-deltak}, where $\Delta k \equiv k - k_c$ is the deviation from the critical value where the catastrophe occurs \eqref{eq:beta-deltak}.  The scalar tilt is $1 - n_s \sim 3/\mathcal{N}_*$ \eqref{tilt}, slightly redder than the observed value.  The tilt increases with increasing $\Delta k$, but cannot be brought into the observed range in these examples.  As with any small-field model the tensor power is negligible.  Since the inflationary trajectories are approximately single field, no significant non-Gaussianity will be generated during the slow-roll phase (although a light transverse field, which is present in some of our examples, could still generate it at reheating). Lastly, the density perturbations depend on the overall scale of the potential, which in turn depends exponentially on the volume of the 4-cycles that produce the relevant terms in the potential \eqref{Vfull}.  Of the two examples we present here, in the $N=5$ case the volume is large and the resulting scale of inflation and perturbation amplitude are very low.  Our $N=8$ example has a higher scale and perturbations that are lower than the observed values by a factor of roughly 7.

If we include phases, the parameter space is much larger.  As we will see, in our $N=5$ example, given almost any non-zero phase, $\Delta k$ can be chosen so that the tilt matches the observed value.  However the inflationary scale is very low in this example, making the amplitude of perturbations far too small.  In the $N=8$ example we have verified that including non-zero phases allows the perturbation amplitude to agree with observation, but the tilt of the spectrum remains slightly too red.  We leave a more systematic investigation of this landscape to future work.

%%%%%%%%%%%%%%%%%%%%%%%%%%%%%%%%%%%%%%%%%%%%%%%%%%%%%%%%%%%
\subsection{\texorpdfstring{$N=5$}{N=5} example} \label{sec:5d_example}
%%%%%%%%%%%%%%%%%%%%%%%%%%%%%%%%%%%%%%%%%%%%%%%%%%%%%%%%%%%
We now present an explicit example of the catastrophe mechanism in a geometry with $N=5$ axions. The geometric data specifying this example is given in Appendix~\ref{app:geometry-n5}. We begin the analysis by setting all phases to zero.  Under the volume rescaling in \eqref{TauScaling}, the potential has two minima when $k < k_c$, and just one minimum when $k > k_c$. 
The charge matrix for this example, ordered by increasing $q\cdot\tau$ at $k=1$, is
\begin{equation} \label{eq:5d_charge_matrix}
  \mathbfcal{Q} =
  \begin{pmatrix}
    1&0&0&0&0\\
    0&0&0&0&1\\
    0&1&0&0&0\\
    2&2&-2&1&1\\
    2&1&-1&1&0\\
    0&0&0&1&0\\
    0&0&1&0&0\\
    7&5&-3&3&2
  \end{pmatrix} \,,
  \qquad
  q\cdot\tau =
  \left(
    6,\ 6.25,\ 24,\ 26,\ 31.875,\ 32,\ 36.125,\ 162.125
  \right) \,.
\end{equation}
The first four rows of the charge matrix have the smallest values of $q \cdot \tau$.\footnote{The raw CYTools output contains two identical copies of the charge vector
\((2,1,-1,1,0)\), both with~\(q\cdot\tau=31.875\). This repeated charge vector is listed only
once in Eq.~\eqref{eq:5d_charge_matrix}.} Recall that the coefficients of the cosines in the potential \eqref{VP} scale as $\Lambda_I^4 \sim e^{-2 \pi k q\cdot\tau}$, so these are the dominant terms in the potential by a very large margin (the ratio $\Lambda_5^4/\Lambda_4^4 = 1.14 \times 10^{-16}$ at $k=1$).  Physically, this means that the fields in the four-dimensional subspace spanned by the first four rows have very large mass compared to the single remaining light direction.  Therefore, to a  good approximation we can set the arguments of the cosines corresponding to each of these first four rows to zero, as they will be minimized on a time scale very short compared to the scale set by the last direction.  

This gives four constraints on the five axion directions. The remaining unfixed direction is the one-dimensional null space of these four charge vectors, parameterized by
\begin{equation}
    2 \pi\,\theta_{\rm{light}} = (0, 0, 1, 2, 0) \vartheta \,,
\end{equation}
where $\vartheta$ is the remaining one-dimensional axion coordinate. Along this direction the first four cosine terms remain at their minima, and the next two terms give a potential in the light direction (the remaining terms are relatively negligible). These terms correspond to the charge vectors
\begin{equation}
  q_1=(2,1,-1,1,0) \,, \qquad q_2=(0,0,0,1,0) \,,
\end{equation}
which have 
\begin{equation}
  2\pi\,q_1\cdot \theta_{\rm light}=\vartheta \,, \qquad
  2\pi\,q_2\cdot \theta_{\rm light}=2\vartheta \,.
\end{equation}
Their respective $q_i\cdot \tau$'s are very close, meaning their coefficients are comparable at least at relatively small volume.

From the potential \eqref{VP}, these two charge vectors give the following contribution to the one-dimensional potential:
\begin{align} \label{eq:V_light}
  V_{\rm light}(\vartheta;k)
  &=
  A_1(k)\left(1-\cos\vartheta\right)
  +A_2(k)\left(1-\cos 2\vartheta\right) \,, \\
  A_1(k)&\propto \,{\frac{255}{8}} k\,
      \exp\left(-\frac{255}{4}\pi k\,\right) \,, \\
  A_2(k)&\propto 32 k \, \exp\left(-64\pi k\right) \,.
\end{align}
Factoring out the first coefficient gives the normalized one-dimensional potential
\begin{equation} \label{1DpotentialN5}
  V_{\rm light}(\vartheta;k)
  \simeq A(k)\left[1-\cos\vartheta+a(k)\left(1-\cos 2\vartheta\right)\right] \,, ~
  a(k)= \frac{32}{255/8}\exp\left[-2\pi k\left(32-\frac{255}{8}\right)\right] \,, 
\end{equation}
and $A(k) \approx 4 \times 10^{-89}$ at $k = 1$ (after putting back in the prefactor from \eqref{VP}). 
This is precisely the first toy model of  \S\ref{Toy model} with the coefficient $a(k)$ now determined by the
Calabi--Yau divisor volumes.  Define
\begin{equation}
  a(k)=\frac{1}{4}-\gamma(k) \,.
\end{equation}
Since $a(k)$ decreases with $k$, the critical value is reached when $\gamma(k_c)=0$, or equivalently
\begin{equation}
  a(k_c)= \frac{1}{4}
  \quad\Rightarrow\quad
  k_c= \frac{4}{\pi}\log \frac{1024}{255} \approx 1.77 \,.
\end{equation}
For $k<k_c$, the potential contains a local minimum and two neighboring critical points. As $k$ approaches the critical value, these points coalesce into a cusp catastrophe. For $k > k_c$, they have disappeared, leaving a shallow hilltop region that can support slow-roll evolution. The number of efolds can be made arbitrarily large by tuning $k$ closer and closer to $k_c$. The conditions for computational control described in \S\ref{stringsec} are satisfied: at $k=k_c$, the smallest divisor volume is $\mathcal{O}(10)$, and the smallest curve volume is approximately $1.33$. 

\begin{figure}[p]
    \centering
    \includegraphics[width=0.75\linewidth]{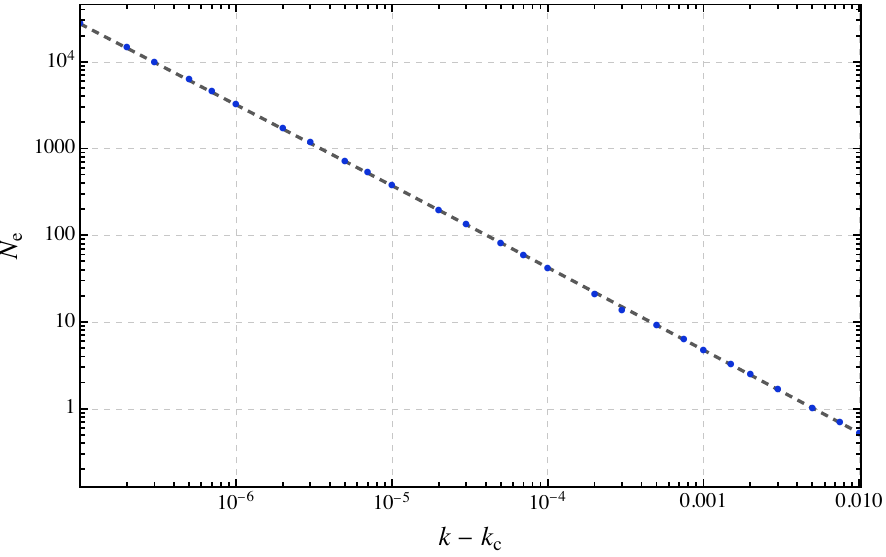}
    \caption{Total number of efolds as a function of the detuning $\Delta k = k-k_c$ for the $N=5$ example. The initial condition is fixed at a canonical field-space distance $\Delta\phi = 10^{-8}M_{\rm Pl}$ from the hilltop. Blue points are numerical trajectories in the potential \eqref{eq:V_light}, while the dashed curve is the analytic estimate from \eqref{eq:efolds-quartic-deltak}, with $\beta_1$ and $c_4$ extracted from \eqref{eq:V_light} at $k=k_c$.}
    \label{fig:2787_N_vs_tuning}
\end{figure}

The analytic hilltop estimate of \eqref{eq:efolds-quartic-deltak} gives a good approximation to the detuning dependence of the efold count obtained from the full numerical trajectories. This comparison is shown in Fig.~\ref{fig:2787_N_vs_tuning} for a fixed initial canonical displacement $\Delta\phi=10^{-8}M_{\rm Pl}$ from the near-critical point.  As expected, the number of efolds grows rapidly as $k$ is tuned toward $k_c$ from above.\footnote{Many efolds of slow-roll inflation can also be achieved for $k < k_c$ so long as the starting point for the field is not in the basin of attraction of the local minimum.}

\begin{figure}[p]
    \centering
    \includegraphics[width=1\linewidth]{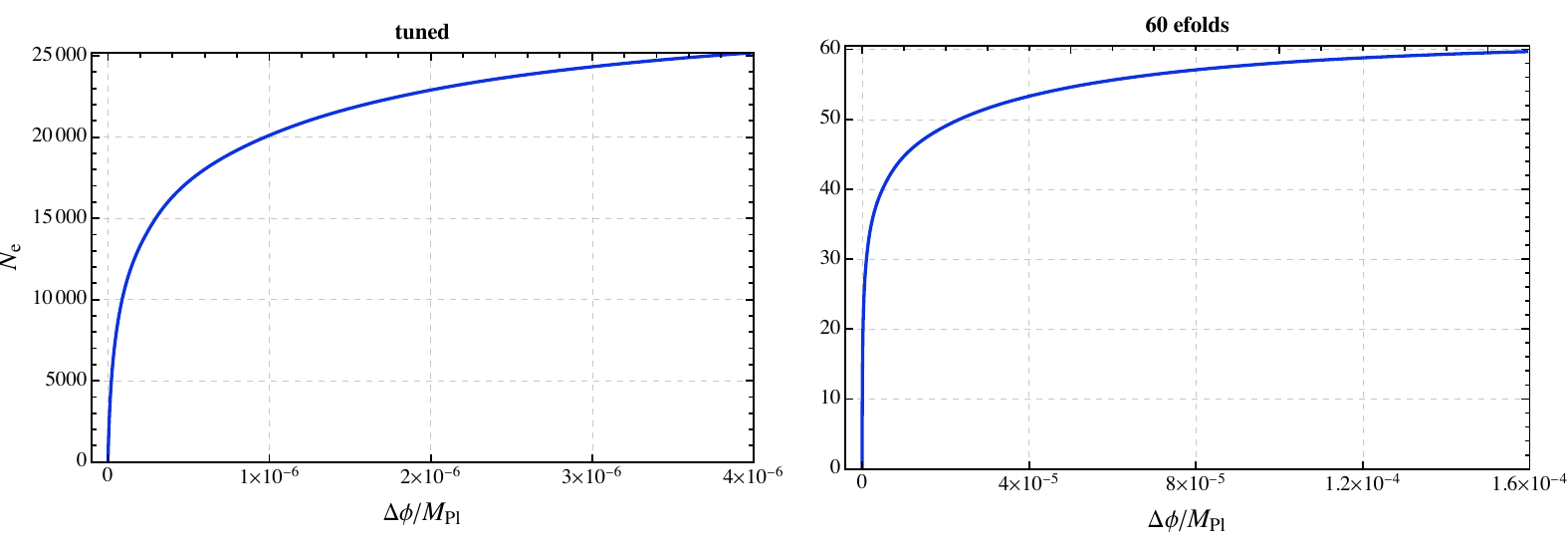}
    \caption{Accumulated number of efolds as a function of the canonically normalized field distance along the gradient flow trajectory for two choices of the volume-scaling parameter, $k$. \textbf{Left}: the finely-tuned case, $k -k_c = 1 \times 10^{-7}$, which produces $\mathcal{N}_e \sim 27349$ efolds before inflation ends. \textbf{Right}: the trajectory chosen to give 60 efolds of inflation, with $k-k_c = 6.77 \times 10^{-5}$. Inflation is terminated when either $\varepsilon>1$ or
    $|\eta_\parallel|>1$. Note that the two panels use different scales.}
    \label{fig:2787_two_panel_efolds}
\end{figure}

Fig.~\ref{fig:2787_two_panel_efolds} explicitly shows the accumulation of efolds as a function of field distance for two representative trajectories: one which is tuned to be very close to the catastrophe point $(k - k_c =  10^{-7})$ which gives 27349 efolds and one much farther away from the catastrophe $(k-k_c = 6.77 \times 10^{-5})$,  chosen to produce 60 efolds.

\begin{figure}[p]
    \centering
    \includegraphics[width=1\linewidth]{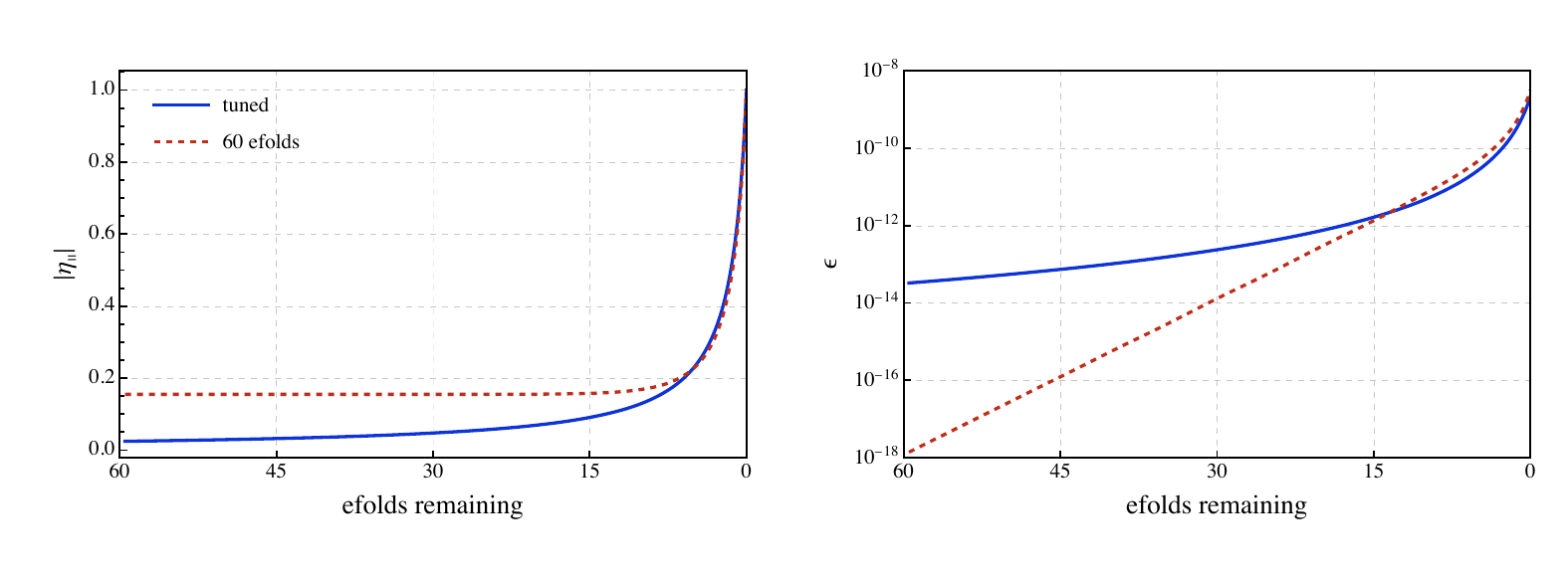}
    \caption{Slow-roll parameters over the final 60 efolds of inflation for the tuned and detuned trajectories. \textbf{Left}: $|\eta_\parallel|$ along the trajectory. \textbf{Right}: $\varepsilon$ along the trajectory. The horizontal axis is the number of efolds remaining before the end of inflation, so evolution proceeds from left to right. The tuned trajectory uses $k -k_c = 1 \times 10^{-7}$, while the 60-efold trajectory uses $k-k_c = 6.77 \times 10^{-5}$.}
    \label{fig:2787_slow_roll_params}
\end{figure}

Although the total number of efolds is highly sensitive to the tuning of $k$, the slow-roll behavior over the final 60 efolds is qualitatively similar. 
This is evident in Fig.~\ref{fig:2787_slow_roll_params}, where we plot slow-roll parameters $\eta_\parallel$ and $\varepsilon$ along the trajectory for the tuned and 60-efold trajectories.
In both cases, $\varepsilon$ remains extremely small and $\eta_\parallel$ grows toward order one and triggers the end of slow-roll, which is expected in small-field inflation.

\begin{figure}[p]
    \centering
    \includegraphics[width=1\linewidth]{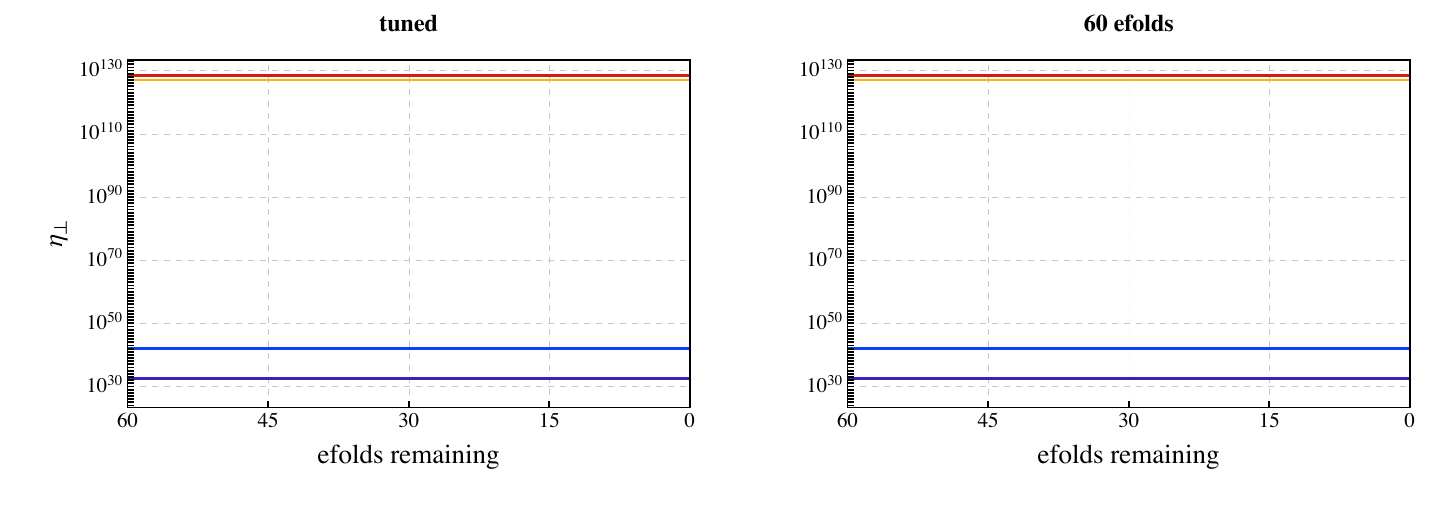}
    \caption{Transverse curvature parameters $\eta_{\perp,i}$ over the final 60 efolds for the tuned (left panel) and 60-efold (right panel) trajectories with the same $k$ values from Fig.~\ref{fig:2787_two_panel_efolds}. The $\eta_{\perp,i}$ are the eigenvalues of the canonical Hessian projected onto the subspace orthogonal to the trajectory. The enormous values of $\eta_{\perp}$ are due to the very small value of $V$ along the inflationary trajectory.  The horizontal axis gives the number of efolds remaining before the end of inflation, so evolution proceeds from left to right.}
    \label{fig:2787_eta_perp}
\end{figure}

We project the canonical Hessian onto the four-dimensional subspace orthogonal to the trajectory and compute the corresponding eigenvalues to get $\eta_{\perp,i}$, which are plotted as a function of efolds remaining in Fig.~\ref{fig:2787_eta_perp}. All four transverse eigenvalues remain positive for both the tuned and the detuned trajectories.
Large positive values of $\eta_{\perp,i}$ correspond to steep transverse
curvature, so in those directions the trajectory lies along a narrow valley of the potential. The absence of negative eigenvalues indicates that there is no tachyonic instability. Thus all orthogonal directions are stabilized against small fluctuations around the trajectory. 

\begin{figure}[p]
    \centering
    \includegraphics[width=0.75\linewidth]{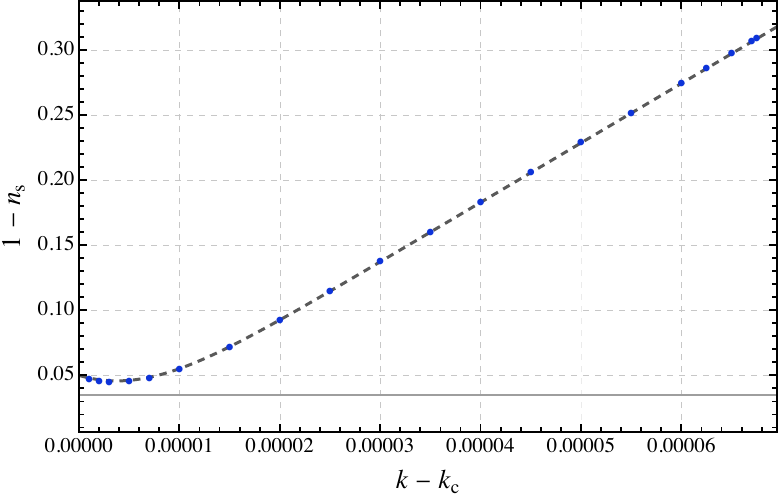}
    \caption{Scalar tilt as a function of detuning from the near-critical value of the volume-scaling parameter. For each trajectory with $\mathcal{N}_e > 60$, we compute $1-n_s$ at 60 efolds before the end of inflation. The blue points are obtained from the full numerical trajectories, the dashed curve is the quartic-hilltop estimate from \eqref{eq:scalar_tilt_analytic}, and the solid gray line is the observed value, $1-n_s = 0.035$ \cite{Planck:2018jri}.}
    \label{fig:2787_scalar_tilt}
\end{figure}

We also examine how the scalar tilt $n_s$ varies with the tuning. For trajectories which have more than $60$ efolds, we evaluate $1- n_s$ using the single-field slow-roll expression
\begin{equation}
    1 - n_s =6 \varepsilon -  2 \eta_\parallel 
\end{equation}
at 60 efolds before the end of inflation. The value of $1-n_s$ increases as $k$ is detuned away from $k_c$, which is shown in Fig.~\ref{fig:2787_scalar_tilt}. The value of $1 - n_s$ from this example is slightly larger than the observed value, so even the highly tuned trajectories are slightly too red to match observations. This depends only on the shape of the potential and is unaffected by its overall scale.

The density perturbation amplitude, which depends on the overall scale, is much farther off from observations. For the two trajectories shown in Fig.~\ref{fig:2787_two_panel_efolds}, $\delta_H \approx 6 \times 10^{-71} $ for the 60-efold trajectory, and $\delta_H \approx 5 \times 10^{-73}$ for the tuned trajectory (the observed value is $\delta_H \approx 1.9 \times 10^{-5}$ \cite{Planck:2018vyg}).  These extremely small values follow from our assumption that $V(\thetavec= \boldsymbol{0}) = 0$ in \eqref{truncpot}. We will see in the following section that the $N=8$ example comes much closer to the observed $\delta_H$.

\begin{figure}[p]
    \centering
\includegraphics[width=0.75\linewidth]{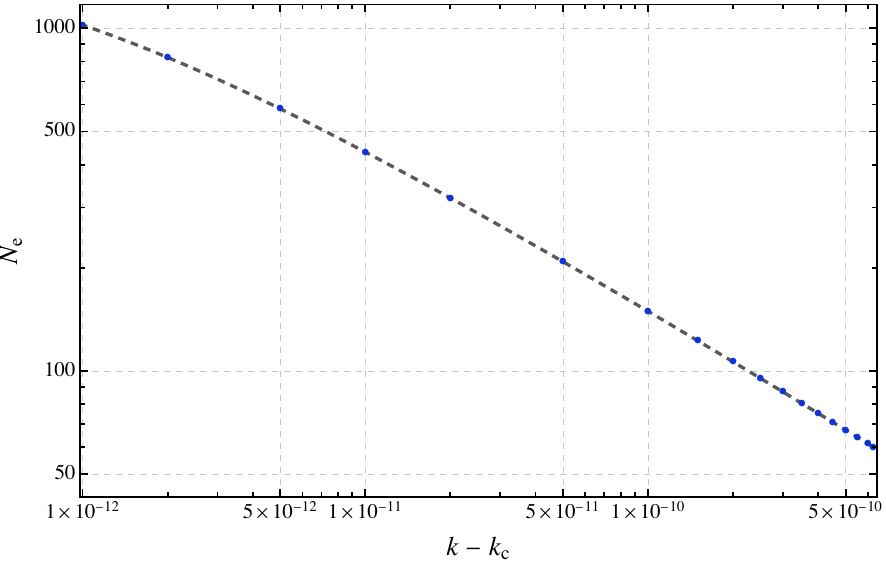}
    \caption{Total number of efolds as a function of the detuning $\Delta k = k-k_c$ for the $N=5$ example with nonzero relative phase $\delta = \pi/4$. The initial condition is fixed at a canonical field-space distance $\Delta\phi = 10^{-8}M_{\rm Pl}$ from the inflection point. Blue points are numerical trajectories, while the dashed curve is the analytic estimate from \eqref{eq:cubic-linear-efolds}, with parameters extracted from the local reduced potential at the fold. The cubic coefficient determines $F$, and the detuning coefficient $\lambda_1$, where $\lambda(k) = \lambda_1 \Delta k$ is obtained from the $k$-derivative at the fold.}
    \label{fig:2787_cubic_efolds}
\end{figure}

\paragraph{Nonzero phases} As we showed in \S\ref{Toy model}, adding a phase to the second cosine term in \eqref{eq:V_light} breaks the symmetry responsible for the quartic cusp. The catastrophe is instead unfolded into a cubic shoulder, which can also be tuned to give many efolds of inflation.  We illustrate this unfolding here for a particular choice of phase, $\delta = \pi/4$, though the same qualitative structure is exhibited for almost any nonzero phase ($\delta > 10^{-8}$). The tuning required to obtain a trajectory with 60 efolds of inflation is more severe for the fold than for the quartic cusp. Near the fold $\mathcal{N}_{e,\text{fold}} \sim 1/\sqrt{\Delta k}$, compared with  $\mathcal{N}_{e,\text{cusp}} \sim 1/\Delta k$ near the cusp. This behavior is evident in Fig.~\ref{fig:2787_cubic_efolds}, where we plot the number of efolds as a function of $\Delta k$.  

\begin{figure}[t!]
    \centering
\includegraphics[width=0.75\linewidth]{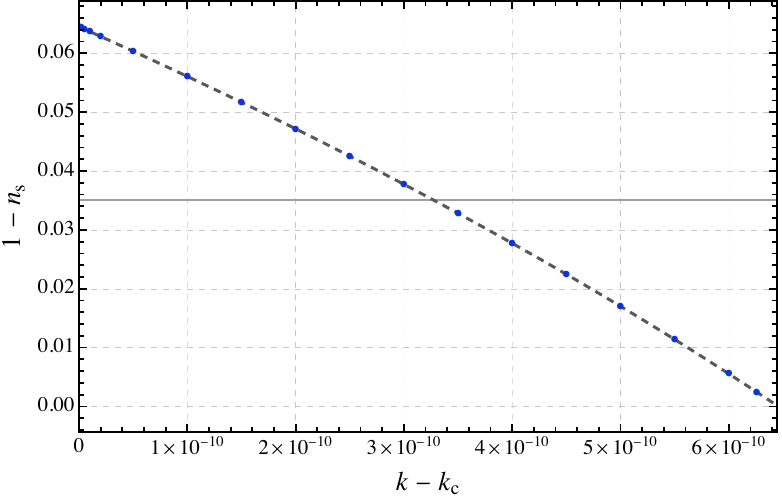}
    \caption{Scalar tilt as a function of detuning from the near-critical value of the volume-scaling parameter for the $N=5$ example with nonzero phase $\delta = \pi/4$. For each trajectory with $\mathcal{N}_e > 60$, we compute $1-n_s$ at 60 efolds before the end of inflation. The blue points are obtained from the full numerical trajectories, the dashed curve is the cubic-linear estimate from \eqref{eq:cubic-linear-tilt}, and the solid gray line is a representative value in the observed range, $1-n_s = 0.035$.}
    \label{fig:2787_cubic_tilt}
\end{figure}

The scalar tilt is no longer tied to the quartic hilltop value from \eqref{eq:scalar_tilt_analytic}, which is too red to match the observed value of $1-n_s$. Instead, by detuning slightly from the fold, the tilt is governed by \eqref{eq:cubic-linear-tilt}, which \textit{can} match the observed value, as shown in Fig.~\ref{fig:2787_cubic_tilt}.

\clearpage

%%%%%%%%%%%%%%%%%%%%%%%%%%%%%%%%%%%%%%%%%%%%%%%%%%%%%%%%%%%
\subsection{\texorpdfstring{$N=8$}{N=8} example}\label{sec:8d_example}
%%%%%%%%%%%%%%%%%%%%%%%%%%%%%%%%%%%%%%%%%%%%%%%%%%%%%%%%%%%
We now present another example of the catastrophe mechanism in a geometry with $N = 8$ axions. Again, we first analyze the case of vanishing phases.
Unlike the previous example, we were not able to analytically reduce it to a one-dimensional potential by fixing heavy terms. Nevertheless, we will see that the inflationary trajectories are approximately linear in the field space, and a one-dimensional approximation describes the results with reasonable accuracy, especially when the volume is tuned very close to the catastrophe. 

The geometric data specifying this example is given in Appendix~\ref{app:geometry}. For $k < k_c$, the potential has five distinct minima, while for $k > k_c$, only one minimum remains. We find that the critical value is $k_c \approx 0.67$. As mentioned in \S\ref{stringsec} this is below the tip of the stretched K\"{a}hler cone, but we have verified that all the curves with volumes less than $1$ at this point are nilpotent (see \S\ref{stringsec}). All potent curves have volumes larger than $1$. Additionally, the smallest divisor volume at this point is $\mathcal{O}(10)$. We therefore conclude that the conditions for control are satisfied.

\begin{figure}[p]
    \centering
    \includegraphics[width=0.75\linewidth]{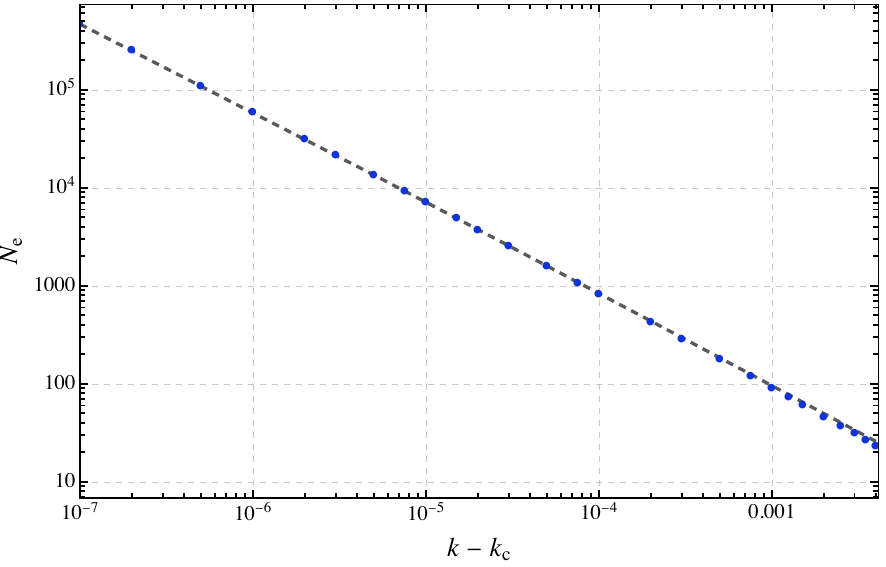}
    \caption{Total number of efolds as a function of detuning from the near-critical value \mbox{$k_c= 0.674506370003365$} of the volume-scaling parameter, with the initial condition fixed at a canonical field-space distance $\Delta\phi = 10^{-8} M_{\rm Pl}$ from the near-critical point. Blue points are numerical trajectories in the full $8$D field space, while the dashed curve is the analytic estimate from \eqref{eq:efolds-quartic-deltak}, with $\beta_1$ and $c_4$ fit to the potential along the trajectory.}
    \label{fig:EfoldsvsTuning}
\end{figure}

\begin{figure}[p]
    \centering
    \includegraphics[width=1\linewidth]{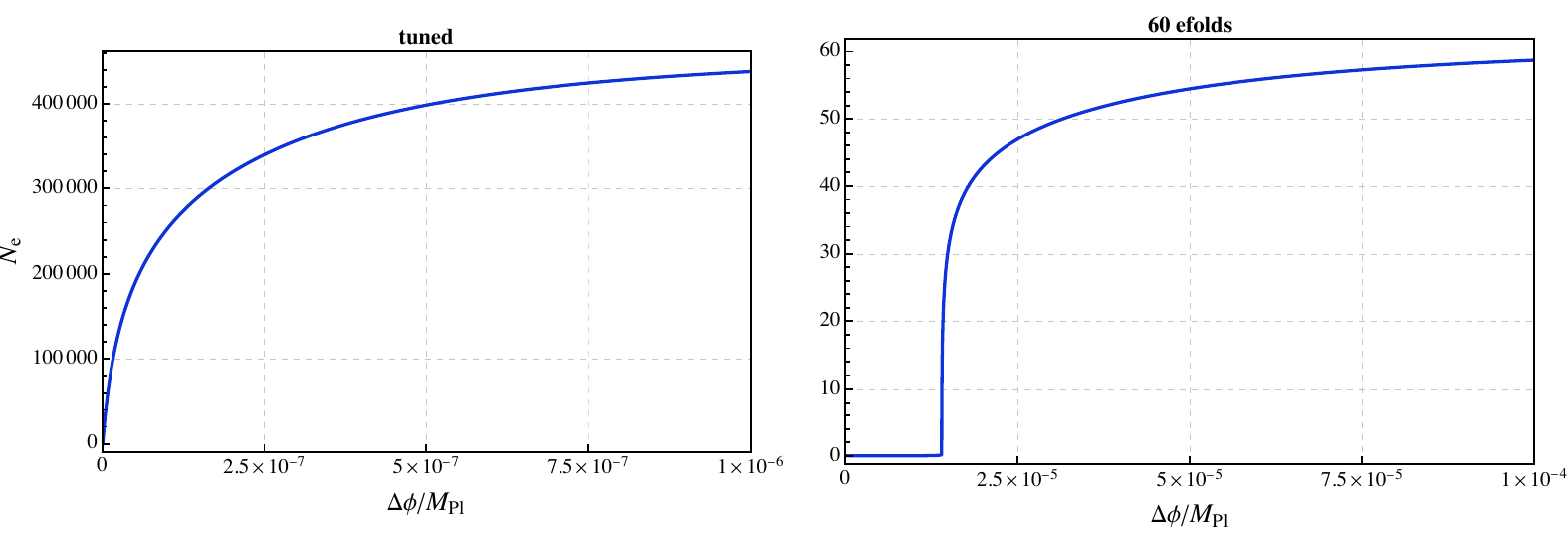}
    \caption{Accumulated number of efolds as a function of the canonically normalized field distance along the gradient flow trajectory for two choices of the volume-scaling parameter, $k$. \textbf{Left}: the finely-tuned case, $k -k_c = 1 \times 10^{-7}$, which produces $\mathcal{N}_e \sim 463115$ efolds before inflation ends. \textbf{Right}: a trajectory chosen to give  60 efolds of inflation, with $k-k_c = 1.5 \times 10^{-3}$. Inflation is terminated when 
    $|\eta_\parallel| = 1$ ($\varepsilon \ll 1$ throughout). Both trajectories begin at the same field-space point, which is not in the inflating region for the detuned case on the right.  
    Note that the two panels use different scales.} 
    \label{fig:efoldComparison}
 \end{figure}

As for the previous example, we plot the total number of efolds as a function of the distance from $k_c$, shown in Fig.~\ref{fig:EfoldsvsTuning}. We show the accumulation of efolds along the gradient descent trajectory for two particular choices of $k$ in Fig.~\ref{fig:efoldComparison}, one very tuned trajectory which produces $\mathcal{N}_e \sim 463115$ efolds, and one trajectory which is chosen so that it  produces $\mathcal{N}_e \sim 60$ efolds.

\begin{figure}[p]
    \centering
    \includegraphics[width=1\linewidth]{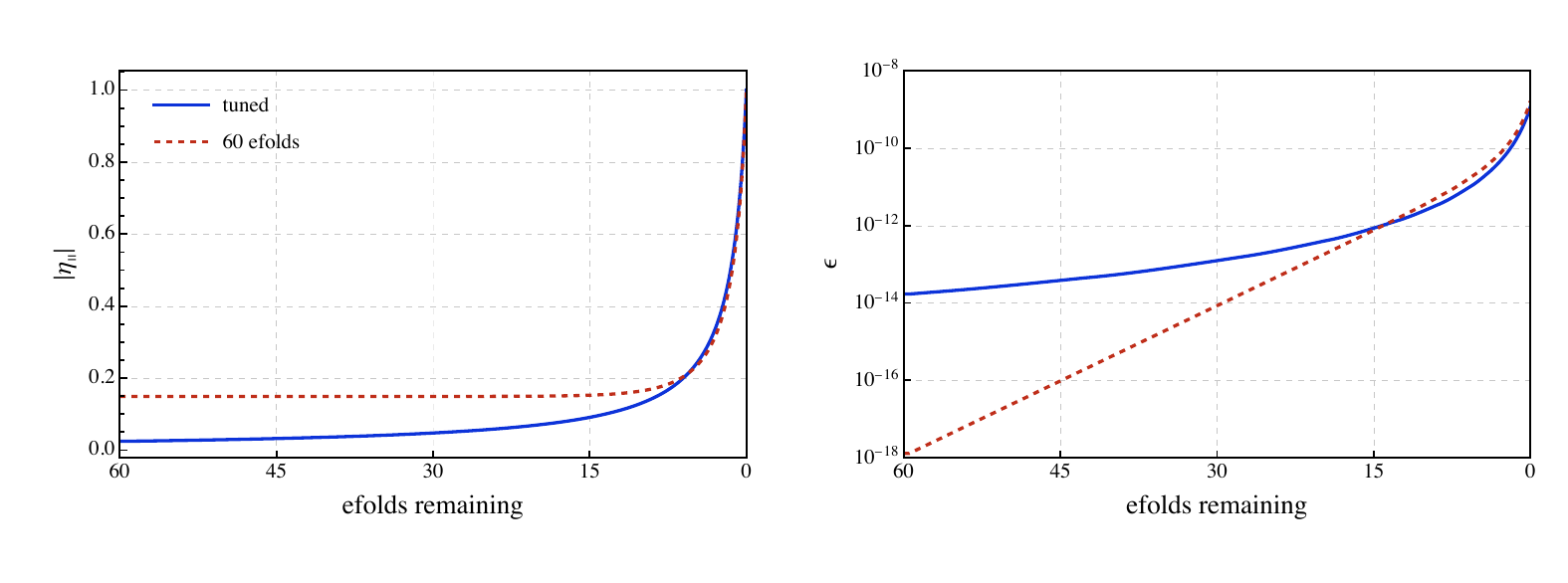}
    \caption{Slow-roll parameters over the final 60 efolds of inflation for the tuned and 60-efold
    trajectories. \textbf{Left}: $|\eta_\parallel|$ along the trajectory. \textbf{Right}: $\varepsilon$ along the trajectory. The horizontal axis is the number of efolds remaining before the end of inflation, so evolution proceeds from left to right. The tuned trajectory uses $k - k_c = 10^{-7}$, while the 60-efold trajectory uses $k - k_c = 1.5 \times 10^{-3}$.} \label{fig:slowRollParams}
\end{figure}

The slow-roll parameters shown in Fig.~\ref{fig:slowRollParams} have very similar behavior to those of the $N = 5$ example, which reflects the fact that both trajectories are controlled by the same local cusp structure.   

\begin{figure}[p]
    \centering
    \includegraphics[width=0.75\linewidth]{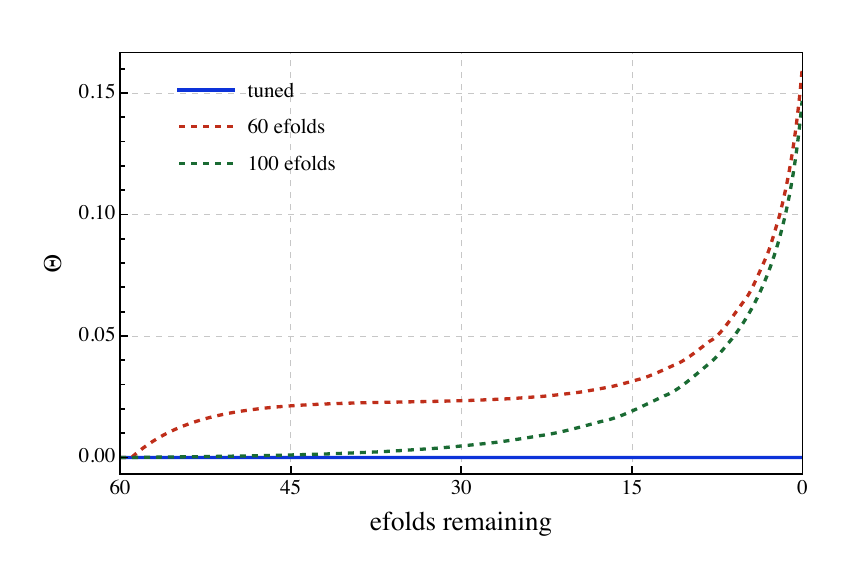}
    \caption{Cumulative turning angle over the final 60 efolds for the tuned (blue) and the 60-efold (dashed red) trajectories from Fig.~\ref{fig:efoldComparison}. The green dashed line is the cumulative turning angle for a trajectory which was tuned such that it has 100 efolds in total ($k - k_c \approx 9 \times 10^{-4}$).  $\Theta$ denotes the accumulated change in the unit tangent vector in canonically normalized field space, defined in \eqref{turning_angle}.}
    \label{fig:CumulativeTurning}
\end{figure}

\begin{figure}[p]
    \centering
    \includegraphics[width=1\linewidth]{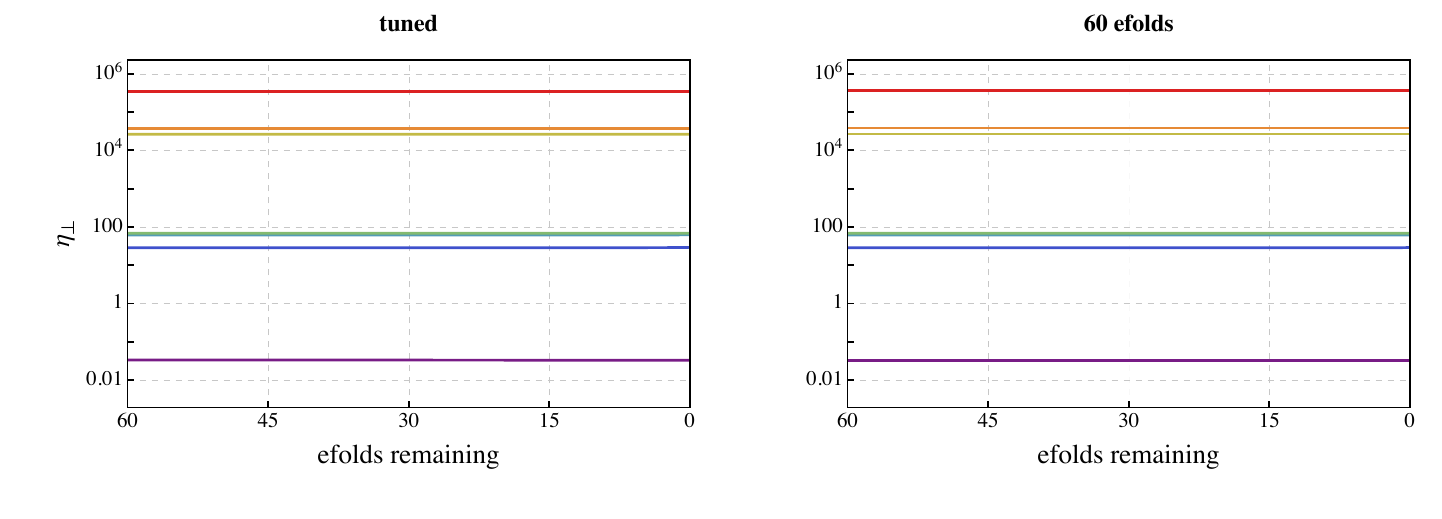}
    \caption{Transverse curvature parameters $\eta_{\perp,i}$ over the final 60 efolds for the tuned (left) and 60-efold (right) trajectories from Fig.~\ref{fig:efoldComparison}. The $\eta_{\perp,i}$ are the eigenvalues of the canonical Hessian projected onto the subspace orthogonal to the trajectory. The horizontal axis gives the number of efolds remaining before the end of inflation, so evolution proceeds from left to right. There is a second light mode in this example in addition to the inflaton. This could lead to significant isocurvature contributions and/or non-Gaussianity, but we have not examined this in detail.}
    \label{fig:eta_perp}
\end{figure}

We quantify the bending of the trajectory by computing the cumulative turning angle in canonically normalized field space. At each point along the trajectory, we define the unit tangent vector by normalizing the field-space velocity,
\begin{equation}
\hat T(s) = \frac{\di \boldsymbol{\phi}/\di s}{|\di \boldsymbol{\phi}/\di s|} \,.
\end{equation}
We then sample the trajectory at closely spaced points and compute the angle between neighboring tangent vectors,
\begin{equation}   
\Delta \Theta_i =
\cos^{-1}\!\left(\hat T_i \cdot \hat T_{i+1}\right) \,.
\end{equation}
The cumulative turn angle at the $j$th point is obtained by summing these increments along the path,
\begin{equation} \label{turning_angle}
\Theta_j = \sum_{i=1}^{j} \Delta \Theta_i \,.
\end{equation}
This quantity measures the total amount by which the trajectory bends in field space during the final 60 efolds of inflation, and is plotted for three different trajectories in Fig.~\ref{fig:CumulativeTurning}. The farther away the $k$ value is from $k_c$, the larger a total turn angle the trajectory accumulates during this period. The tuned example has a cumulative turning angle which is negligible. As in the $N=5$ example, in Fig.~\ref{fig:eta_perp} we plot the slow-roll parameters $\eta_{\perp,i}$ in the subspace orthogonal to the trajectory. We identify a second light mode, which suggests that there may be a two-dimensional plateau where the conditions for inflation are satisfied, although we have not examined this in detail.

\begin{figure}[p]
    \centering
    \includegraphics[width=0.75\linewidth]{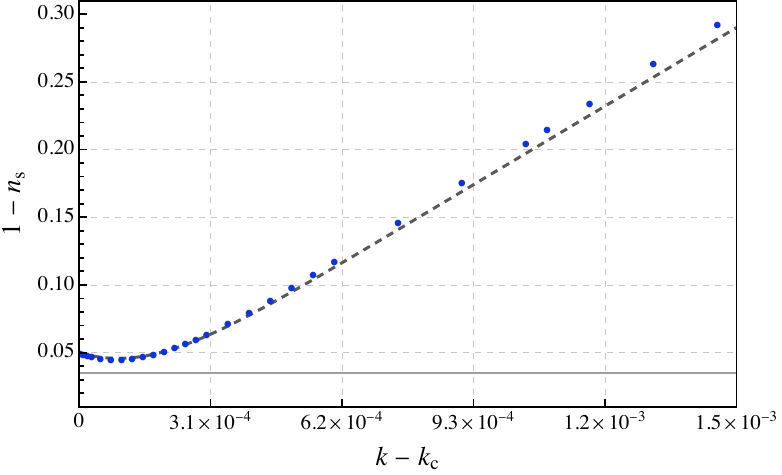}
    \caption{Scalar tilt as a function of detuning from the near-critical value of the volume-scaling parameter. For each trajectory with $\mathcal{N}_e > 60$, we compute $1-n_s$ at 60 efolds before the end of inflation. The blue points are obtained from the full numerical trajectories, the dashed curve is the quartic-hilltop estimate from \eqref{eq:scalar_tilt_analytic}, and the solid gray line is the observed value, $1-n_s = 0.035$ \cite{Planck:2018jri}.}
    \label{fig:ns_vs_tuning}
\end{figure}

We show the scalar tilt as a function of tuning for this example in Fig.~\ref{fig:ns_vs_tuning}. The result is qualitatively and numerically similar to the previous example: $1-n_s$ is minimized for trajectories closest to the catastrophe and increases as $k$ is detuned away from $k_c$, so the spectrum becomes progressively redder. This similarity reflects the fact that both examples are controlled by the same local quartic-cusp structure over the observable window. Unlike in the previous example, here we see that the hilltop estimate in Appendix~\ref{infparams} starts to slightly deviate from the numerical simulations as $k$ gets more detuned, which reflects the fact that this potential is not exactly one-dimensional.  

The density perturbation for this example is much closer to the observed value than the $N = 5$ example. At 60 efolds from the end of inflation the 60-efold trajectory has $\delta_H \approx 3 \times 10^{-6}$ (roughly 7 times smaller than the observed value), and  the tuned trajectory has  $\delta_H \approx 2 \times 10^{-10}$.

\begin{figure}[p]
    \centering
    \includegraphics[width=0.75\linewidth]{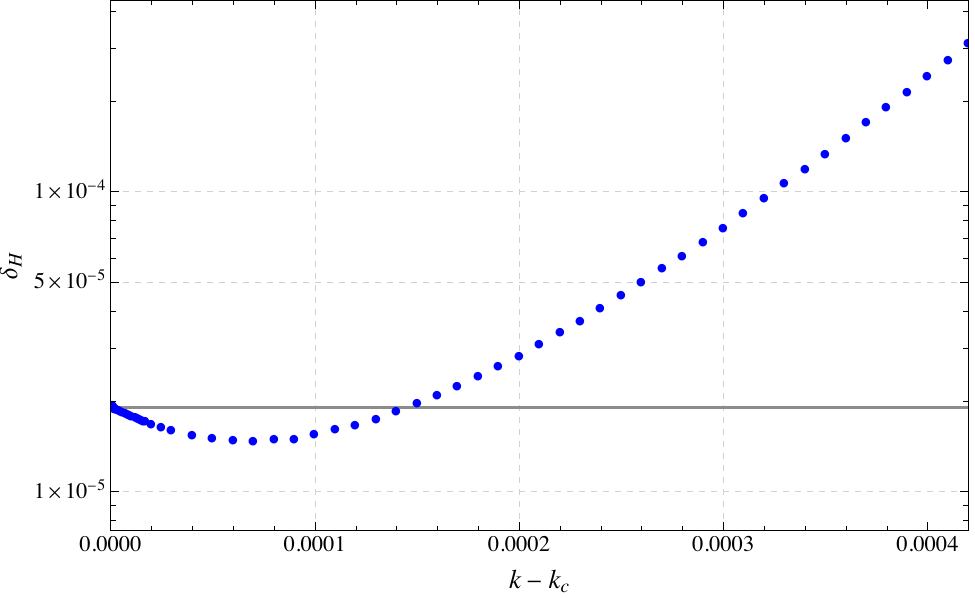}
    \caption{Density perturbation amplitude $\delta_H$ as a function of detuning $\Delta k$ for the $N=8$ example with a phase $\delta= 0.04$. Each blue point is computed from a full eight-field numerical trajectory and evaluated at 60 efolds before the end of inflation. The solid gray line shows the observed value $\delta_H \simeq 1.9 \times 10^{-5}$ \cite{Planck:2018vyg}.}
    \label{fig:102_phase_deltaH}
\end{figure}

\begin{figure}[p]
    \centering
    \includegraphics[width=1\linewidth]{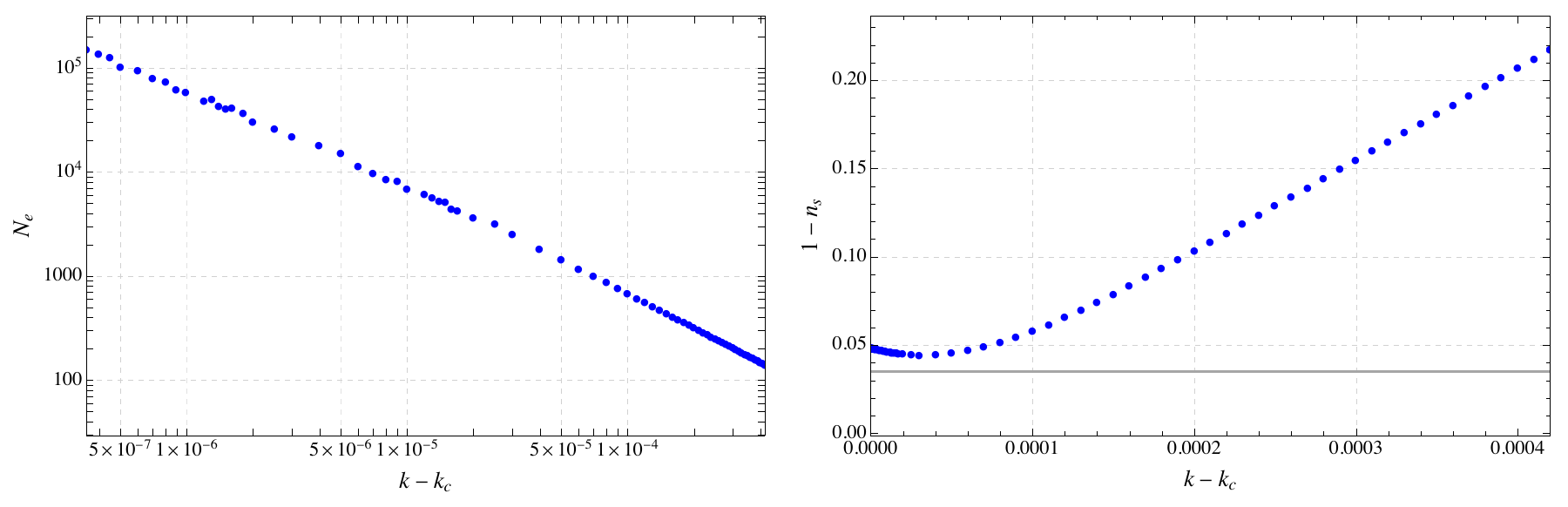}
    \caption{Inflationary observables as a function of the detuning $\Delta k$ for the case with non-zero phase. \textbf{Left}: total number of efolds $\mathcal{N}_e$  obtained from the numerical trajectories. \textbf{Right}: scalar tilt, $1-n_s$, evaluated 60 efolds before the end of inflation for trajectories with $\mathcal{N}_e > 60$.}
    \label{fig:102_phase_efolds_tilt}
\end{figure}

\paragraph{Nonzero phases} The value of $\delta_H$ can be increased enough to match the observed value by introducing nonzero phases in this example. We add a phase $\delta = 0.04$ to the term in the potential corresponding to the second row of the charge matrix in Table~\ref{tab:dominant-charges}. The location of the catastrophe changes to $k_c \simeq 0.5$. In Fig.~\ref{fig:102_phase_deltaH}, we show that for two different values of $k$, the numerical trajectory has the observed value of $\delta_H$. The number of efolds and scalar tilt as a function of detuning are shown in Fig.~\ref{fig:102_phase_efolds_tilt}.

\clearpage
%%%%%%%%%%%%%%%%%%%%%%%%%%%%%%%%%%%%%%%%%%%%%%%%%%%%%%%%%%%
\section{Conclusions}
%%%%%%%%%%%%%%%%%%%%%%%%%%%%%%%%%%%%%%%%%%%%%%%%%%%%%%%%%%%
Our findings suggest that axion fields in string theory compactifications can give rise to small-field inflation via a mechanism based on catastrophes. The key to achieving this is to carefully tune the volumes of the Calabi--Yau manifold's cycles, which allows for a near-cancellation of terms in the axion potential. This fine-tuning creates flat regions in the potential.  This mechanism relies on the assumption that the saxions can be stabilized at a point where this fine-tuning occurs, a crucial open question that requires further investigation. The dynamics of the full system, including the kinetic terms and the coupling between the axions and saxions, must be fully understood to determine if a slow-roll inflationary trajectory of this type is indeed possible. 

As with any small-field inflation model, fine-tuning the potential is not sufficient.  First, the expectation value of the inflaton field $\langle \phi \rangle$ must be tuned to lie in or evolve to the (small-field) inflationary plateau of the potential,  with sufficiently small time derivative that it does not overshoot it before slow-roll can begin.  Moreover, East et al.~\cite{East:2015ggf} discovered that if the  spatial variation of the field on a spatial slice exceeds the field-space width of the inflationary plateau, $\delta\phi \gtrsim \Delta\phi$, some spatial regions sample the steep part of the potential and pull the average field off the inflationary plateau before inflation can begin.  This criterion, and its implications for the fragility of small-field inflation, was explored further by Clough et al.~\cite{Clough:2016ymm}. Note that tunneling from a false vacuum can simultaneously produce highly homogeneous initial conditions, prevent ``overshoot'' of the inflationary plateau due to the spatial curvature, and help explain any tuning necessary to achieve a sufficient number of efolds of inflation to inflate away the spatial curvature and allow structure to form \cite{Freivogel:2005vv}.  It would be interesting to investigate this possibility further in this context.

It is worth commenting on the relation of this work to the various swampland conjectures.  
The distance conjecture~\cite{Ooguri:2006in} asserts that field excursions
$\Delta\phi\gtrsim M_{\rm Pl}$ should be accompanied by a tower of  exponentially light states, presumably invalidating the effective field theory description.
This is an obstacle for axion alignment and $N$-flation constructions
that rely on super-Planckian effective decay constants.  Our mechanism avoids this entirely, as the  inflationary trajectory is confined to $\Delta\phi\ll f\ll M_{\rm Pl}$. 
The (refined) de~Sitter conjecture~\cite{Obied:2018sgi,Garg:2018reu,Ooguri:2018wrx}
requires that a controlled potential satisfy either $M_{\rm Pl}|\nabla V|\ge c\,V$ or
$\min (\nabla^2 V)\le -c'\,V/M_{\rm Pl}^2$, with $c,c'\sim\mathcal{O}(1)$.  In small-field models matching the observed red tilt
fixes $|\eta_\parallel|\simeq (1-n_s)/2 \approx0.02$ at horizon exit.  Moreover, as the field approaches the catastrophe both
$|\nabla V|/V$ and $|\min (\nabla^2 V)|/V$ tend to zero, so a neighborhood of the
cusp violates both inequalities for any fixed $c'$.  Therefore compatibility with the refined de~Sitter conjecture holds only for  $c'\lesssim0.02$ and only if $\Delta k$ cannot be tuned to arbitrarily small values.
The axionic weak gravity conjecture~\cite{ArkaniHamed:2006dz} requires, in its simplest form, 
$f S_{\rm inst}\lesssim M_{\rm Pl}$.  This is marginally satisfied here: for $f/M_{\rm Pl}\sim10^{-2}$ and
$S_{\rm inst}=2\pi q\cdot\tau\sim\mathcal O(10^2)$,  one has
$fS_{\rm inst}/M_{\rm Pl}\sim\mathcal{O}(1)$.  In a multi-axion setting the more stringent condition is
the convex-hull form~\cite{Heidenreich:2015nta}, which we have not investigated in detail.

It should be noted that the examples of slow-roll inflation studied in this work involve models with modest numbers of axions ($N=5$ and $N=8$). We expect models with larger $N$ in general to require more tuning: the slow-roll parameters scale inversely with the size of the kinetic terms, which are known to decrease with $N$ \cite{Mehta:2021pwf, Demirtas:2018akl}.

A number of other open questions remain.  In the zero-phase case ($\delta_I=0$ in \eqref{VP}) all the examples we have found are of the cusp, rather than fold, type (i.e.~three critical points merge, resulting in a catastrophe where both second and third derivatives vanish, rather than two critical points merging).  In our examples this is due to a discrete symmetry, but we leave open the question of whether all examples are of this type.  Generic phase assignments break the symmetry and can produce fold catastrophes, with inflationary potentials of the cubic shoulder type, but we have not systematically investigated the full parameter space of arbitrary phase assignments even in the examples we have identified so far.  It would be interesting to perform a more comprehensive numerical study of geometries to better understand the set of potential inflationary points in the axion sector.

\paragraph{Acknowledgements} We would like to thank Joan La Madrid, Viraf Mehta, Richard Nally, and Sonia Paban for useful discussions. The work of MK and CN is supported by NSF grant PHY-2412899. The work of NG is supported in part by a grant from the Simons Foundation (602883,
CV) and the DellaPietra Foundation.

%%%%%%%%%%%%%%%%%%%%%%%%%%%%%%%%%%%%%%%%%%%%%%%%%%%%%%%%%%%
\appendix
%%%%%%%%%%%%%%%%%%%%%%%%%%%%%%%%%%%%%%%%%%%%%%%%%%%%%%%%%%%

%%%%%%%%%%%%%%%%%%%%%%%%%%%%%%%%%%%%%%%%%%%%%%%%%%%%%%%%%%%
\section{Inflation near cusp and fold catastrophes}\label{infparams}
%%%%%%%%%%%%%%%%%%%%%%%%%%%%%%%%%%%%%%%%%%%%%%%%%%%%%%%%%%%

%%%%%%%%%%%%%%%%%%%%%%%%%%%%%%%%%%%%%%%%%%%%%%%%%%%%%%%%%%%
\subsection{Quartic hilltop}
%%%%%%%%%%%%%%%%%%%%%%%%%%%%%%%%%%%%%%%%%%%%%%%%%%%%%%%%%%%
For the quartic hilltop toy model (near a cusp catastrophe, where three critical points merge)
\begin{equation}
V(\phi)=V_0\left(
1-\frac{\beta}{2}x^2-\frac{c_4}{4}x^4
\right) \,, 
\qquad
x\equiv \frac{\phi}{f} \,,
\qquad
F\equiv \frac{f}{M_{\rm Pl}} \,,
\end{equation}
the slow-roll parameters are, to leading order near the saddle,
\begin{equation}
\varepsilon_V \simeq \frac{1}{2F^2}\left(\beta x+c_4x^3\right)^2 \,,
\qquad
\eta_V \simeq -\frac{1}{F^2}\left(\beta+3c_4x^2\right) \,.
\end{equation}
Let $\mathcal N_*$ denote the number of efolds remaining when the CMB scale
exits the horizon.  In the regime $x_*\ll x_{\rm end}$, the efold integral gives
\begin{equation}
\mathcal N_*
\simeq
\frac{F^2}{2\beta}
\log\!\left(
1+\frac{\beta}{c_4x_*^2}
\right) \,.
\end{equation}
Equivalently, defining
\begin{equation}
\rho\equiv\frac{\beta}{F^2} \,,
\end{equation}
one finds
\begin{equation}
x_*^2
\simeq
\frac{\beta}{c_4\left(e^{2\rho\mathcal N_*}-1\right)} \,.
\end{equation}
The scalar tilt is therefore
\begin{equation}
\label{eq:scalar_tilt_analytic}
n_s-1
\simeq
2\eta_{V,*}-6\varepsilon_{V,*}
\simeq
-2\rho\left(
1+\frac{3}{e^{2\rho\mathcal N_*}-1}
\right) \,,
\end{equation}
where the contribution from $\varepsilon_{V,*}$ is negligible for $F\ll1$.

In the pure quartic limit, $\beta\to0$, this reduces to
\begin{equation}
x_*^2 \simeq \frac{F^2}{2c_4\mathcal N_*} \,,
\qquad
\eta_{V,*} \simeq -\frac{3}{2\mathcal N_*} \,,
\end{equation}
and hence
\begin{equation} \label{tilt}
n_s \simeq 1-\frac{3}{\mathcal N_*} \,.
\end{equation}
For $\mathcal N_*=60$, this gives $n_s\simeq0.95$.  For positive $\beta$,
the tilt initially becomes slightly less red, but the maximum of the analytic
expression is only $n_s\simeq0.954$ for $\mathcal N_*=60$.

The density-perturbation amplitude may be written as
\begin{equation}
\delta_H
\simeq
\frac{1}{5\sqrt{3}\pi M_{\rm Pl}^3}
\frac{V^{3/2}}{|V'|} \,,
\end{equation}
giving
\begin{equation}
\delta_H
\simeq
\frac{1}{5\sqrt{3}\pi}
\frac{\sqrt{V_0}}{M_{\rm Pl}^2}
\frac{F}{x_*\left(\beta+c_4x_*^2\right)} \,.
\end{equation}
In the pure quartic limit this simplifies to
\begin{equation}
\delta_H
\simeq
\frac{(2\mathcal N_*)^{3/2}}{5\sqrt{3}\pi}
\frac{\sqrt{c_4V_0}}{f^2} \,.
\end{equation}

%%%%%%%%%%%%%%%%%%%%%%%%%%%%%%%%%%%%%%%%%%%%%%%%%%%%%%%%%%%
\subsection{Cubic shoulder}
%%%%%%%%%%%%%%%%%%%%%%%%%%%%%%%%%%%%%%%%%%%%%%%%%%%%%%%%%%%
For the cubic shoulder model (near a fold catastrophe, where two critical points merge), write
\begin{equation}
V(\phi)=V_0\left(1-\lambda x-x^3\right) \,,
\qquad
x\equiv\frac{\phi}{f} \,,
\qquad
F\equiv\frac{f}{M_{\rm Pl}} \,,
\qquad
\lambda\ge0 \,.
\label{eq:cubic-linear-inflation}
\end{equation}
The field orientation is chosen so that it rolls toward increasing $x$.  In
the small-field regime $|x|\ll1$, $V\simeq V_0$ and
\begin{equation}
\varepsilon_V
\simeq
\frac{1}{2F^2}\left(\lambda+3x^2\right)^2 \,,
\qquad
\eta_V
\simeq
-\frac{6x}{F^2} \,.
\label{eq:cubic-slow-roll}
\end{equation}
Slow-roll ends at $\eta_V\simeq-1$, so
\begin{equation}
x_{\rm end}\simeq\frac{F^2}{6} \,.
\label{eq:cubic-xend-leading}
\end{equation}
For the exactly cubic potential, retaining the $1-x^3$ denominator gives the
exact equation
\begin{equation}
F^2x_{\rm end}^3+6x_{\rm end}-F^2=0 \,,
\end{equation}
whose real root is
\begin{equation}
x_{\rm end}
=
\sqrt[3]{\frac12+\sqrt{\frac14+\frac8{F^6}}}
+
\sqrt[3]{\frac12-\sqrt{\frac14+\frac8{F^6}}}
=
\frac{F^2}{6}\left( 1-\frac{F^6}{216}+\mathcal O(F^{12}) \right) \,.
\end{equation}

\paragraph{Exactly cubic inflation} For $\lambda=0$, the efold integral gives
\begin{equation}
\mathcal N_*
\simeq
\frac{F^2}{3}\left(\frac1{x_*}-\frac1{x_{\rm end}}\right)
\simeq
\frac{F^2}{3x_*}-2 \,,
\end{equation}
and therefore
\begin{equation}
x_*
\simeq
\frac{F^2}{3(\mathcal N_*+2)}
=
\frac{F^2}{3\mathcal N_*}\left( 1-\frac2{\mathcal N_*}
+\mathcal O(\mathcal N_*^{-2})\right) \,.
\end{equation}
Substitution into \eqref{eq:cubic-slow-roll} yields
\begin{align}
n_s
&\simeq
1-\frac{4}{\mathcal N_*+2}
=
1-\frac4{\mathcal N_*}+\mathcal O(\mathcal N_*^{-2}) \,,
\label{eq:pure-cubic-tilt}\\
r
&\equiv16\varepsilon_{V,*}
\simeq
\frac{8F^6}{9(\mathcal N_*+2)^4}
=
\frac{8F^6}{9\mathcal N_*^4}
\left( 1-\frac8{\mathcal N_*}+\mathcal O(\mathcal N_*^{-2})\right) \,.
\label{eq:pure-cubic-r}
\end{align}
Thus tensors are negligible, while the exactly cubic model is too red:
$n_s\simeq0.935$ for $\mathcal N_*=60$.

The scalar amplitude is
\begin{equation}
A_s
\simeq
\frac{V_0}{24\pi^2M_{\rm Pl}^4\varepsilon_{V,*}}
\simeq
\frac{3}{4\pi^2}
\frac{V_0}{M_{\rm Pl}^4}
\frac{(\mathcal N_*+2)^4}{F^6} \,,
\label{eq:pure-cubic-As}
\end{equation}
or equivalently
\begin{equation}
\delta_H=\frac25\sqrt{A_s}
\simeq
\frac{\sqrt3}{5\pi}
\frac{\sqrt{V_0}}{M_{\rm Pl}^2}
\frac{(\mathcal N_*+2)^2}{F^3} \,.
\end{equation}

\paragraph{Cubic plus linear inflation}
The linear term competes with the cubic contribution to $V'$ when
$\lambda\sim x_*^2\sim F^4/\mathcal N_*^2$.  Define
\begin{equation}
\mu^2\equiv\frac{3\lambda}{F^4} \,.
\label{eq:mu-cubic-linear}
\end{equation}
Using $x_{\rm end}\simeq F^2/6$, the efold integral becomes
\begin{align}
\mathcal N_*
&\simeq
F^2\int_{x_*}^{x_{\rm end}}
\frac{\di x}{\lambda+3x^2}
\nonumber\\
&=
\frac1\mu\left[
\tan^{-1}\!\left(\frac1{2\mu}\right)
-
\tan^{-1}\!\left(\frac{3x_*}{\mu F^2}\right)
\right] \,.
\label{eq:cubic-linear-efolds}
\end{align}
Inverting,
\begin{equation}
\frac{3x_*}{\mu F^2}
=
\tan\!\left[
\tan^{-1}\!\left(\frac1{2\mu}\right)-\mu\mathcal N_*
\right] \,.
\label{eq:cubic-linear-xstar-exact}
\end{equation}
For large $\mathcal N_*$, set
\begin{equation}
s\equiv\mu\mathcal N_* \,.
\end{equation}
Then
\begin{equation}
x_*
=
\frac{F^2}{3\mathcal N_*}\,s\cot s
+\mathcal O\!\left(\frac{F^2}{\mathcal N_*^2}\right) \,.
\label{eq:cubic-linear-xstar}
\end{equation}
The available slow-roll interval disappears as $s\to\pi$, so obtaining
$\mathcal N_*$ efolds requires, parametrically, $\mu\lesssim\pi/\mathcal N_*$.  The
red-tilted branch relevant for observation has $0<s<\pi/2$.

Using \eqref{eq:cubic-linear-xstar}, one obtains
\begin{align}
n_s
&\simeq
1-\frac{4s\cot s}{\mathcal N_*} \,,
\label{eq:cubic-linear-tilt}\\
r
&\simeq
\frac{8F^6}{9\mathcal N_*^4}
\left(\frac{s}{\sin s}\right)^4\ll1 \,,
\label{eq:cubic-linear-r}\\
A_s
&\simeq
\frac{3}{4\pi^2}
\frac{V_0}{M_{\rm Pl}^4}
\frac{\mathcal N_*^4}{F^6}
\left(\frac{\sin s}{s}\right)^4 \,,
\label{eq:cubic-linear-As}\\
\delta_H
&\simeq
\frac{\sqrt3}{5\pi}
\frac{\sqrt{V_0}}{M_{\rm Pl}^2}
\frac{\mathcal N_*^2}{F^3}
\left(\frac{\sin s}{s}\right)^2 \,.
\label{eq:cubic-linear-deltaH}
\end{align}
The detuning required for a specified $s$ is
\begin{equation}
\lambda
=
\frac{F^4s^2}{3\mathcal N_*^2} \,.
\label{eq:cubic-linear-lambda-tuning}
\end{equation}
If the observationally allowed interval is $n_s\in[n_1,n_2]$, then
\begin{equation}
(1-n_2)\frac{\mathcal N_*}{4}
<
s\cot s
<
(1-n_1)\frac{\mathcal N_*}{4} \,.
\end{equation}
For example, $s=3\sqrt3/5$ gives $n_s\simeq0.959$ at
$\mathcal N_*=60$ and
\begin{equation}
\lambda
=
\frac9{25}\frac{F^4}{\mathcal N_*^2} \,.
\end{equation}
Thus a small displacement from the fold, of order
$\lambda\sim F^4/\mathcal N_*^2$, can move the scalar tilt from the overly red
exact-cubic value into the observed range while leaving the tensor signal
negligible.

%%%%%%%%%%%%%%%%%%%%%%%%%%%%%%%%%%%%%%%%%%%%%%%%%%%%%%%%%%%
\section{Finding critical points of the axion potential}
\label{sec:critical_points_algorithm}
%%%%%%%%%%%%%%%%%%%%%%%%%%%%%%%%%%%%%%%%%%%%%%%%%%%%%%%%%%%
Here we review the numerical procedure used to find critical points of the axion potential. The method follows \cite{Gendler:2023kjt}, and exploits two special features of the potentials that arise in the KS axiverse: the instanton coefficients are highly hierarchical, and the charge matrices are sparse. These facts make it possible to reduce the search for critical points from an $N$-dimensional transcendental problem to a much lower-dimensional one.

We begin with the truncated potential
\begin{equation} \label{truncpot}
    V_P(\thetavec)
    =
    \sum_{I=1}^{P}
    \Lambda_I^4\left[1-
    \cos\left(2\pi(\mathbfcal Q\thetavec)^I + \delta_I \right)\right] \,,
\end{equation}
where the retained terms are chosen by ordering the instanton contributions by amplitude and discarding terms that are sufficiently subdominant  to the leading terms. The phases $\delta_I$ are set by the phase of the flux superpotential $W_0$, as well as by one-loop Pfaffians, whose computation is at present out of reach (see \cite{Kim:2023cbh, Alexandrov:2022mmy} for progress in this regard).  The $(1 - \cos)$ structure is equivalent to adding a constant such that when all the phases $\delta_I = 0$, the global minimum is at $\thetavec = \boldsymbol{0}$ and has zero vacuum energy $V(\thetavec= \boldsymbol{0})=0$. For notational reasons we will set the phases to zero for the rest of this Appendix, and experiment with restoring them in the specific examples in \S\ref{resultssec}.

The first $N$ linearly independent charge vectors are collected into an invertible matrix, which we denote by $\Qtilde$. By construction these terms give the dominant contribution to all $N$ axion directions. Additional retained terms, whose charges are linearly dependent on the rows of $\Qtilde$ but whose amplitudes are larger than a fixed threshold, are collected into a matrix $\Qbar$. Thus
\begin{equation}
    V_P(\thetavec)
    =
    \sum_{i=1}^{N}
    \Lambdatilde_i^4 \left[ 1-\cos\left(2\pi(\Qtilde\thetavec)^i\right) \right ]
    +
    \sum_{a=1}^{P-N}
    \Lambdabar_a^4 \left[ 1-
\cos\left(2\pi(\Qbar\thetavec)^a\right)\right] \,.
\end{equation}
The truncation threshold is chosen so that terms much smaller than the dominant terms on which they depend are discarded. For sparse, order-one charges, such subleading terms are not expected to create qualitatively new critical points, although they can split degeneracies associated with an enlarged fundamental domain \cite{Gendler:2023kjt}.

It is useful to change coordinates to variables in which the dominant terms are diagonal,
\begin{equation}
    \thetatildevec
    =
    \Qtilde\thetavec \,.
\end{equation}
In these coordinates the potential becomes
\begin{equation}
    V_P(\thetatildevec)
    =
    \sum_{i=1}^{N}
    \Lambdatilde_i^4
    \left[1-\cos(2\pi\widetilde{\theta}^i)\right]
    +
    \sum_{a=1}^{P-N}
    \Lambdabar_a^4
    \left[ 1- \cos\left(2\pi\alphavec^{(a)}\cdot\thetatildevec\right)\right],
    \qquad
    \alphavec^{(a)}
    =
    (\Qtilde^{-1})^{\top}\qbar^{(a)} \,.
\end{equation}
The transformed charges $\alphavec^{(a)}$ are typically sparse. Consequently, many axions appear only in the first, separable sum. For any direction $i$ with $\alpha_i^{(a)}=0$ for all retained subleading terms $a$, the critical point equation is solved analytically:
\begin{equation}
    \sin(2\pi\widetilde{\theta}^i)=0 \,, 
    \qquad
    \widetilde{\theta}^i=0 ~\text{or}~  \frac12 \,.
\end{equation}
These directions therefore need not be included in the numerical root search. The remaining directions $j$ obey
\begin{equation}
    \sin(2\pi\widetilde{\theta}^j)
    +
    \sum_{a=1}^{P-N}
    \frac{\Lambdabar_a^4}{\Lambdatilde_j^4}
    \alpha_j^{(a)}
    \sin\left(2\pi\alphavec^{(a)}\cdot\thetatildevec\right)
    =
    0 \,.
\end{equation}
We solve this reduced system numerically inside a fundamental domain of the truncated potential. The fundamental domain is determined by the lattice of discrete shifts that leave all retained cosine arguments invariant; equivalently, it is the intersection of the hyperplane $\{ \mathbfcal Q\thetavec \, | \, \thetavec \}$ with the integer lattice. In practice this lattice basis can be computed using the Smith normal form, as described in \cite{Gendler:2023kjt}. Initial points are sampled uniformly in this reduced fundamental domain, and any solution returned outside the domain is shifted back by the appropriate lattice vector.

Once the critical points have been found, we classify them by the Hessian. Because the amplitudes $\Lambdatilde_i^4$ can differ by many orders of magnitude, it is numerically advantageous to rescale the Hessian as
\begin{equation}
    \widehat H_{ij}
    =
    \frac{H_{ij}}
    {\sqrt{|\Lambdatilde_i^4|}\sqrt{|\Lambdatilde_j^4|}} \,.
\end{equation}
This rescaling preserves the number of positive and negative Hessian eigenvalues, but removes the large hierarchies from the matrix entries.  Critical points with positive definite Hessian are minima.

%%%%%%%%%%%%%%%%%%%%%%%%%%%%%%%%%%%%%%%%%%%%%%%%%%%%%%%%%%%
\section{Geometric data for the \texorpdfstring{$N=5$}{N=5} example}
\label{app:geometry-n5}
%%%%%%%%%%%%%%%%%%%%%%%%%%%%%%%%%%%%%%%%%%%%%%%%%%%%%%%%%%%
The geometry used in \S\ref{sec:5d_example} is the Calabi--Yau hypersurface associated with the
four-dimensional reflexive polytope \(\Delta^\circ\) with vertices
\begin{equation}\label{eq:vert}
\begin{aligned}
\mathrm{Vert}(\Delta^\circ)=\{&
(1,0,0,0),\,
(-1,2,-1,-1),\,
(0,0,0,1),\,
(-2,-1,0,0),\\
&
(-1,-2,2,0),\,
(0,0,1,0),\,
(-2,0,-1,0),\,
(0,1,0,0)
\} \,.
\end{aligned}
\end{equation}
For this choice of lattice, the Hodge numbers are
\begin{equation}
        h^{1,1}=5 \,,
    \qquad
    h^{2,1}=75 \,,
    \qquad
    \chi=-140 \,,
\end{equation}
so the axion sector contains \(N=h^{1,1}=5\) axions. 
Let \(v_1,\ldots,v_8\) denote the vertices in the order of \eqref{eq:vert},
and order the point configuration as
\begin{equation}
p_0=(0,0,0,0),\qquad
p_i=v_i\ \ (i=1,\ldots,8),\qquad
p_9=(0,-1,1,0).
\end{equation}
The fine, regular, star triangulation used in our analysis is the
regular triangulation induced, in this ordering, by the height vector
\begin{equation}
\mathbf{h}=(0,\,3,\,0,\,0,\,0,\,0,\,-2,\,0,\,-1,\,1).
\end{equation}
We choose the point at the tip of the stretched K\"ahler cone, normalized so that the smallest curve volume is one. At this point, the total Calabi--Yau volume is
\begin{equation}
    \mathcal V = 149.395833333 \,.
\end{equation}
The divisor volumes in the GLSM basis of toric divisors are
\begin{equation}
    \tau = \left(6,\;24,\; \frac{289}{8},\;32,\; \frac{25}{4}\right) \,,
\end{equation}
and the curve volumes range from 1 to 12.5. 

The instanton charges are listed in \eqref{eq:5d_charge_matrix}. The eigenvalues of the axion kinetic matrix are
\begin{equation}
\mathrm{eig}(K_{ij}) =
\big(
3.57\times 10^{-3},
2.20\times 10^{-4},
1.06\times 10^{-4},
8.05\times 10^{-5},
2.54\times 10^{-5}
\big) \,.
\end{equation}
These values correspond to sub-Planckian axion decay constants of order
\begin{equation}
    f/M_{\rm Pl} \sim \sqrt{\mathrm{eig}(K_{ij})}
    \sim 10^{-2} \,.
\end{equation}

%%%%%%%%%%%%%%%%%%%%%%%%%%%%%%%%%%%%%%%%%%%%%%%%%%%%%%%%%%%
\section{Geometric data for the \texorpdfstring{$N=8$}{N=8} example}
%%%%%%%%%%%%%%%%%%%%%%%%%%%%%%%%%%%%%%%%%%%%%%%%%%%%%%%%%%%
\label{app:geometry}
The geometry used in \S\ref{sec:8d_example} is the Calabi--Yau hypersurface associated with the
four-dimensional reflexive polytope \(\Delta^\circ\) with vertices
\begin{equation} \label{eq:vert102}
\begin{aligned} 
\mathrm{Vert}(\Delta^\circ)=\{&
(0,0,0,1),\,
(1,0,0,0),\,
(-1,-1,1,0),\,
(-1,1,-1,0),\,
(1,-1,-1,-1),
\\
&
(1,1,1,-1),\,
(0,-1,0,0),\,
(0,0,-1,0),\,
(0,0,1,0),\,
(0,1,0,0)
\} \,.
\end{aligned}
\end{equation}
For this choice of lattice, the Hodge numbers are
\begin{equation}
        h^{1,1}=8 \,,
    \qquad
    h^{2,1}=28 \,,
    \qquad
    \chi=-40 \,.
\end{equation}
Thus the axion sector contains \(N=h^{1,1}=8\) axions. 
Let \(v_1,\ldots,v_{10}\) denote the vertices in the order of \eqref{eq:vert102},
and order the point configuration as
\begin{equation}
\begin{aligned}
p_0&=(0,0,0,0),\qquad
p_i=v_i\quad (i=1,\ldots,10),\\
p_{11}&=(-1,0,0,0),\qquad
p_{12}=(1,0,0,-1).
\end{aligned}
\end{equation}
The fine, regular, star triangulation used in our analysis is the
regular triangulation induced, in this ordering, by the height vector
\begin{equation}
\mathbf{h}
=(0,\,11,\,11,\,13,\,13,\,14,\,14,\,11,\,11,\,11,\,11,\,11,\,12).
\end{equation}

We choose the point at the tip of the stretched K\"ahler cone, normalized so that the smallest curve volume is one. At this point, the total Calabi--Yau volume is
\begin{equation}
    \mathcal V = 126 \,.
\end{equation}
The divisor volumes in the GLSM basis of toric divisors are
\begin{equation}
    \tau = (45,\;17,\;17,\;14.5,\;14.5,\;15.5,\;15.5,\;25) \,,
\end{equation}
and the curve volumes range from 1 to 3. 

The truncated axion potential is constructed from the leading instanton charges, ordered by the size of $q_I\cdot \tau$. These charges are listed in Table~\ref{tab:dominant-charges}. The first column gives $q_I\cdot \tau$ at the stretched-cone tip, while the remaining columns give the integer charge vector $q_I$ in the GLSM basis.

\begin{table}[!ht] 
\centering
\[
\begin{array}{c|rrrrrrrr}
q_I\cdot\tau
& \multicolumn{8}{c}{q_I} \\
\hline
14.0 & -1 &  1 &  1 &  0 &  0 &  0 &  0 &  1 \\
14.5 &  0 &  0 &  0 &  1 &  0 &  0 &  0 &  0 \\
14.5 &  0 &  0 &  0 &  0 &  1 &  0 &  0 &  0 \\
15.5 &  0 &  0 &  0 &  0 &  0 &  0 &  1 &  0 \\
15.5 &  0 &  0 &  0 &  0 &  0 &  1 &  0 &  0 \\
15.5 &  0 & -1 &  1 & -1 &  1 &  0 &  1 &  0 \\
15.5 &  0 &  1 & -1 & -1 &  1 &  1 &  0 &  0 \\
16.0 &  1 &  0 &  0 & -1 & -1 &  0 &  0 &  0 \\
17.0 &  0 &  0 &  1 &  0 &  0 &  0 &  0 &  0 \\
17.0 &  0 &  1 &  0 &  0 &  0 &  0 &  0 &  0 \\
25.0 &  0 &  0 &  0 &  0 &  0 &  0 &  0 &  1 \\
45.0 &  1 &  0 &  0 &  0 &  0 &  0 &  0 &  0
\end{array}
\]
\caption{Dominant instanton charges retained in the truncated axion potential
for the $N=8$ example. The first column gives $q_I\cdot\tau$ at the
stretched-cone tip.}
\label{tab:dominant-charges}
\end{table}

At the same point, the eigenvalues of the axion kinetic matrix are
\begin{equation}
\begin{aligned}
\mathrm{eig}(K_{ij}) =
\big(&
8.20\times 10^{-4},
6.35\times 10^{-4},
5.97\times 10^{-4},
3.13\times 10^{-4}, \\
&
1.24\times 10^{-4},
9.15\times 10^{-5},
8.30\times 10^{-5},
5.84\times 10^{-5}
\big) \,.
\end{aligned}
\end{equation}
These values correspond to sub-Planckian axion decay constants of order
\begin{equation}
    f/M_{\rm Pl} 
    \sim 10^{-2} \,.
\end{equation}

%\clearpage
%%%%%%%%%%%%%%%%%%%%%%%%%%%%%%%%%
\bibliographystyle{klebphys2}
\bibliography{refs}
%%%%%%%%%%%%%%%%%%%%%%%%%%%%%%%%%

\end{document}